\documentclass[useAMS,usenatbib]{mn2e}
\usepackage{epsfig}

\title[Simulating seed BH formation]{Simulating the formation of massive seed black holes in the early Universe. II: 
Impact of rate coefficient uncertainties}
\author[S.~C.~O. Glover]{Simon C. O. Glover \\
Universit\"{a}t Heidelberg, Zentrum f\"{u}r Astronomie, Institut f\"{u}r Theoretische Astrophysik, \\ Albert-Ueberle-Stra{\ss}e 2, 
69120 Heidelberg, Germany
}

\begin{document}
\maketitle

\begin{abstract}
We investigate how uncertainties in the chemical and cooling rate coefficients relevant for a metal-free gas
influence our ability to determine the critical ultraviolet field strength required to suppress H$_{2}$ cooling in 
high-redshift atomic cooling halos. The suppression of H$_{2}$ cooling is a necessary prerequisite for the
gas to undergo direct collapse and form an intermediate mass black hole. These black holes can then act
as seeds for the growth of the supermassive black holes (SMBHs) observed at redshifts $z \sim 6$.
The viability of this model for SMBH formation depends on the critical ultraviolet field strength, $J_{\rm crit}$: if 
this is too large, then too few seeds will form to explain the observed number density of SMBHs.
We show in this paper that there are five key chemical reactions whose rate coefficients are uncertain
enough to significantly affect $J_{\rm crit}$. The most important of these is the collisional ionization of
hydrogen by collisions with other hydrogen atoms, as the rate for this process is very poorly constrained
at the low energies relevant for direct collapse. The total uncertainty introduced into $J_{\rm crit}$ by this and
the other four reactions could in the worst case approach a factor of five. 
We also show that the use of outdated or inappropriate values for the rates of some chemical reactions in previous
studies of the direct collapse mechanism may have significantly affected the values of $J_{\rm crit}$
determined by these studies.
\end{abstract}


\section{Introduction}
Observations of quasars at redshifts $z > 6$ demonstrate that supermassive black holes (SMBHs) with
masses of order $10^{9} \: {\rm M_{\odot}}$ had already managed to form by the time that the Universe
was one billion years old. However, this is difficult to understand if these SMBHs form from the seed 
black holes with masses $M \sim 10 \: {\rm M_{\odot}}$ that are produced as the endpoints of the evolution
of massive stars. The problem is one of timescales: even if we assume that accretion onto these seed
black holes is very efficient, the time required for one of them to grow to a mass of $10^{9} \: {\rm M_{\odot}}$
can easily exceed the age of the Universe at $z > 6$ \citep{th09}. 

For this reason, an alternative model for the formation of the progenitors of high-redshift SMBHs has
recently attracted considerable interest. In this model, known as the direct collapse model, gas in a
protogalaxy with a virial temperature $T_{\rm vir} > 10^{4} \: {\rm K}$ that is illuminated by an extremely
strong extragalactic radiation field collapses quasi-isothermally, maintaining its temperature in the range
$T \sim 5000$--10000~K through the combined effects of atomic hydrogen cooling and H$^{-}$ bound-free
cooling \citep{om01}. The high temperature keeps the Jeans mass in the collapsing gas high and inhibits
fragmentation during the collapse \citep[see e.g.][]{bl03,begel06,sbh10}. The precise outcome of the collapse 
remains uncertain, but the most likely outcomes are either the formation of a short-lived supermassive star
or a quasi-star, a massive, optically-thick core whose centre has already collapsed to form a black hole
\citep{begel06,begel10,schleicher13}. 
In either case, the end result is the formation of an intermediate mass black hole (IMBH) with a mass
$M \sim 10^{4} \: {\rm M_{\odot}}$ or larger.

The time required for this much larger seed black hole to grow by accretion into a $10^{9} \: {\rm M_{\odot}}$ 
SMBH is short enough that this mechanism can explain the existence of the observed high redshift SMBHs.
In addition, the strength of the extragalactic radiation field required to suppress H$_{2}$ formation
and the associated cooling in the collapsing gas is much larger than the mean value produced by 
early stellar populations \citep{har00,gb06}. This means that the formation of an IMBH by direct collapse
will be a rare event, occurring only in a small number of protogalaxies located in highly clustered regions
where the local radiation field strength can be orders of magnitude above the mean value  
\citep{dijk08,ahn09,agar12}. The viability of the direct collapse model depends upon just how rare this outcome
is, and hence on how strong the radiation field needs to be in order to sufficiently suppress H$_{2}$ formation.
If direct collapse occurs too rarely, then not enough IMBH progenitors will be produced to explain the
observed number density of SMBHs at high redshift. 

Unfortunately, there is little agreement on the precise value of the radiation field strength required.
This is usually quantified in terms of the specific intensity of the ultraviolet portion of the
radiation field at the Lyman limit. Following \citet{har00}, it is common to measure this in units of
$10^{-21} \: {\rm erg s^{-1} cm^{-2} Hz^{-1} sr^{-1}}$ and to write it as $J_{21}$. The minimum value of 
$J_{21}$ for which H$_2$ cooling is sufficiently suppressed is then commonly denoted as $J_{\rm crit}$.
Physically, the value of $J_{\rm crit}$ depends strongly on the shape of the spectrum of the extragalactic
radiation field \citep{sbh10,soi14,latif15,ak15,agar15} and on the strength of any extragalactic X-ray background or 
cosmic ray background \citep{io11,it14,latif15}. However, our ability to determine $J_{\rm crit}$ accurately from 
simulations is also affected by uncertainties arising from the way in which the chemistry of H$_2$ is modelled 
in these simulations.

There are three main ways in which the chemical model adopted can affect $J_{\rm crit}$. First, its value
in some circumstances is highly sensitive to the way in which the effects of H$_{2}$ self-shielding are
modelled \citep{wh11,soi14}. Second, for reasons of computational efficiency, it is common to model the chemistry 
of the gas using a highly simplified chemical network, but if the model is made too simple, then errors of
up to a factor of a few can be introduced into $J_{\rm crit}$ (\citeauthor{glover15}, 2015; hereafter, Paper I).
Finally, as we explore in this paper, $J_{\rm crit}$ is also sensitive to the values adopted for the rate coefficients
of some of the key chemical reactions occurring in the gas. 

Since most chemical rate coefficients are not known
absolutely precisely, uncertainties in the input rate coefficients inevitably introduce an uncertainty into the
value of $J_{\rm crit}$ predicted by the simulations. In this study, we attempt to quantify the size of this
uncertainty and to identify which chemical reactions make the largest contributions to it. In addition,
as many previous numerical studies of the direct collapse model have adopted values for some of 
the chemical rate coefficients that are now known to differ significantly from the actual values, we also
explore how sensitive $J_{\rm crit}$ is to these choices. 

The structure of our paper is as follows. In Section~\ref{method}, we briefly present the numerical
method we use in our study. In Section~\ref{chemunc}, we quantify the sensitivity of $J_{\rm crit}$ to the
rate of each of the chemical reactions included in our model. Using this information, we identify a
subset of reactions for which the sensitivity is large and critically discuss the uncertainties that exist
in the rate coefficients for each of these reactions. In Section~\ref{coolunc}, we carry out a similar analysis
for the different cooling processes included in our model. In Section~\ref{discuss}, we combine our
results and attempt to estimate the overall uncertainty in $J_{\rm crit}$ due to uncertainties
in the chemical and cooling rate coefficients. We also compare the results of our analysis with those
of previous studies of the direct collapse model. Finally, we summarize the main results of our paper 
in Section~\ref{conc}. 

\section{Numerical method}
\label{method}
\subsection{The one-zone model}
We model the thermal and chemical evolution of gravitationally collapsing gas in halos illuminated by a
strong extragalactic radiation field using the same one-zone model as
described in detail in Paper I. We assume that the gas density evolves as
\begin{equation}
\frac{{\rm d}\rho}{{\rm d}t} = \frac{\rho}{t_{\rm ff}},
\end{equation}
where $t_{\rm ff} = (3\pi / 32 G \rho)^{1/2}$ is the gravitational free-fall time. The thermal
evolution of the gas is followed by solving the internal energy equation 
\begin{equation}
\frac{{\rm d}e}{{\rm d}t} = \frac{p}{\rho^{2}} \frac{{\rm d}\rho}{{\rm d}t} + \Gamma - \Lambda,
\end{equation}
where $e$ is the internal energy density, $p = (\gamma - 1)e$ is the gas pressure, 
$\Gamma$ is the radiative and chemical heating rate per unit volume and $\Lambda$ is the 
radiative and chemical cooling rate per unit volume. These are computed using the detailed
atomic and molecular cooling function introduced in  \citet{Glover09} and updated in Paper I.

To model the chemical evolution of the gas, we use the conservative reduced network derived in 
Paper I. This consists of 25 reactions involving 8 different chemical species: e$^{-}$, H$^{+}$,
H, H$^{-}$, H$_{2}^{+}$, H$_{2}$, He and HeH$^{+}$. In Paper I
we showed that the difference between the value of $J_{\rm crit}$ derived using this network
and that derived using a comprehensive chemical model of primordial gas is around 1\%, much
smaller than the uncertainties considered in this paper. The full list of reactions included in 
our model is given in Table~\ref{tab:react}. 

We model the effects of the external ultraviolet background using two different spectral shapes:
a $10^{5}$~K black-body, cut off at $h\nu > 13.6$~eV to account for absorption by neutral
hydrogen, and a $10^{4}$~K black-body cut off at the same energy. For brevity, we refer to 
these two spectral shapes in the remainder of the paper as T5 and T4 spectra, respectively.
The normalization of the radiation field is specified in terms of its specific intensity at the 
Lyman limit, measured in units of $10^{-21} \: {\rm erg s^{-1} cm^{-2} Hz^{-1} sr^{-1}}$, 
and written as $J_{21}$.

We assume that the gravitationally collapsing gas is optically thin in the continuum and model the effects of H$_{2}$
self-shielding using the modified version of the \citet{db96} shielding function given in 
\citet{whb11}:
\begin{eqnarray}
f_{\rm sh}(N_{\rm H_{2}}, T) & = & \frac{0.965}{(1 + x/b_{5})^{1.1}} + \frac{0.035}{(1+x)^{0.5}} \nonumber \\
& & \times \exp \left[-8.5 \times 10^{-4} (1+x)^{0.5} \right],
\end{eqnarray}
where $x = N_{\rm H_{2}} / 5 \times 10^{14} \: {\rm cm^{-2}}$, $b_{5} = b / 10^{5} \: {\rm cm \: s^{-1}}$,
and $b$ is the Doppler broadening parameter, which we assume to be dominated by the effects
of thermal broadening. To compute $N_{\rm H_{2}}$, we assume that the dominant contribution comes from 
within a central dense core with radius equal to the Jeans length, and hence estimate $N_{\rm H_{2}}$ as
$N_{\rm H_{2}} = n_{\rm H_{2}} \lambda_{\rm J}$, where  $\lambda_{\rm J}$ is the Jeans length.

%In addition to H$_{2}$ self-shielding, we also account for the shielding of H$_{2}$ by the Lyman series
%lines of atomic hydrogen, using the expression \citep{wh11}
%\begin{equation}
%f_{\rm sh, H}(N_{\rm H}) = \frac{1}{(1 + x_{\rm H})^{1.6}} \times \exp (-0.15 \, x_{\rm H}),
%\end{equation}
%where $x_{\rm H} = N_{\rm H} / 2.85 \times 10^{23} \: {\rm cm^{-2}}$. We estimate $N_{\rm H}$
%using the approximation $N_{\rm H} = n_{\rm H} R_{\rm c}$.

As in Paper I, we caution that the simple approximation used here to compute $N_{\rm H_{2}}$
%and $N_{\rm H}$ 
may yield values that are systematically incorrect by up to an order of magnitude.
The absolute values derived here for $J_{\rm crit}$ should therefore be treated with caution. 
However, this systematic uncertainty should not affect the conclusions we draw here regarding
the sensitivity of $J_{\rm crit}$ to uncertainties in the chemical rate coefficient adopted in the model.

\subsection{Initial conditions}
In paper I, we found, in common with previous authors, that the value of $J_{\rm crit}$ depended
only very weakly on our choices for the initial temperature, density and ionization state of the gas.
The reason for this lack of sensitivity is that the question of whether or not enough H$_{2}$ forms
in the gas to provide efficient cooling to temperatures $T \ll 10^{4} \: {\rm K}$ is generally decided
once the gas has already reached a number density of 100--1000~cm$^{-3}$, by which time it
retains only a weak memory of its initial conditions (see e.g.\ \citealt{sbh10} or Paper I).

Therefore, in this study we consider runs performed using only a single set of initial conditions. 
We set the initial temperature to $T_{0} = 8000$~K, the initial hydrogen nuclei number density
to $n_{0} = 0.3 \: {\rm cm^{-3}}$, and assume that the initial fractional ionization and H$_{2}$ fractional 
abundance are close to those found in the IGM prior to the onset of
Pop.\ III star formation ($x_{\rm e, 0} = 2 \times 10^{-4}$, $x_{\rm H_{2},0} = 2 \times 10^{-6}$).
We assume that the gas is initially in charge balance, so that $x_{\rm H^{+}, 0} = x_{\rm e, 0}$,
and place all of the hydrogen not accounted for by H$^{+}$ or H$_{2}$ in atomic form. We also
assume that the helium starts entirely in neutral atomic form, and that the initial abundances
of the other three chemical species in our network (H$^{-}$, H$_{2}^{+}$, and HeH$^{+}$) are
zero. We note that these initial conditions are the same as those adopted in runs 2 and 5 in paper I.

%Starting with these initial conditions, we then run simulations with both a T4 and a T5 spectrum,

\section{Impact of chemical rate coefficient uncertainties on $J_{\rm crit}$}
\label{chemunc}

\subsection{How sensitive is $J_{\rm crit}$ to the rate of each reaction?}
\label{howsense}
As the starting point for our study, we first explore how sensitively our derived value of $J_{\rm crit}$ 
depends on the rate coefficient of each of the reactions in our chemical model. We do this for purely
pragmatical reasons: if the value of $J_{\rm crit}$ is insensitive to the value of a particular rate coefficient,
there is little point in spending time and effort critically assessing how well we know that value of that
rate coefficient. 

To assess how sensitive $J_{\rm crit}$ is to variations in the rate coefficients of the reactions in our
reduced network, we construct 52 different variants of our chemical model.
In each of these variant models, the rate of one of the rate coefficients for our reduced set of 26
reactions is adjusted either upwards or downwards by a factor of $10^{0.5}$ from its fiducial value. 
We then determine $J_{\rm crit}$ for each of these models, for both a T4 and a T5 spectrum. By comparing
the outcome of the case in which a rate coefficient was adjusted upwards with the outcome
of the case in which the same rate coefficient was adjusted downwards, we can quantify the effect of an 
order of magnitude change in that rate coefficient.

The results of this analysis are shown in Table~\ref{tab:react}. The sensitivity values we quote in the table
are defined as the ratio of the largest and the smallest values of $J_{\rm crit}$ that we obtain when modifying 
the rate coefficient of the specified reaction, i.e.\
\[
\mbox{Sensitivity} = \frac{J_{\rm crit, max}}{J_{\rm crit, min}}
\]
With this definition, in a case where $J_{\rm crit}$ depends linearly on the rate coefficient in question we would 
expect to obtain a sensitivity of 10, while a value of 1 shows us that $J_{\rm crit}$ is sensitive only to the inclusion
of that reaction in the network, and not to the value of the rate coefficient adopted for the reaction.

From Table~\ref{tab:react}, we see that $J_{\rm crit}$ is highly sensitive to the values of the rate coefficients 
for only a small subset of the reactions included in our minimal model. Specifically, if we restrict our attention
to those reactions for which the sensitivity exceeds 1.25 (i.e.\ where an order of magnitude change in the rate
coefficient results in a change in $J_{\rm crit}$ of greater than 25\%), then we find that in runs performed with
a T4 spectrum, the high sensitivity reactions are numbers 1-3, 5-8, 10 and 12. In runs performed with a T5
spectrum, we find high sensitivities for a slightly different subset of reactions, numbers 1-3, 5, 6, 8, 10-12, and 22.
This difference in behaviour is a consequence of the difference in the identity of the key process suppressing
H$_{2}$ cooling in the runs with different incident spectra. In runs with a T4 spectrum, H$^{-}$ photodetachment
plays a much greater role than H$_{2}$ photodissociation in suppressing the H$_{2}$ abundance. Therefore,
we find that in these runs, $J_{\rm crit}$ is only weakly sensitive to the H$_{2}$ photodissociation rate (reaction 1),
but has a much greater sensitivity to the H$^{-}$ photodetachment rate (reaction 7). On the other hand, in runs
with a T5 spectrum, H$_{2}$ photodissociation dominates, and so the value of $J_{\rm crit}$ depends strongly
on the rate of reaction 1, but has hardly any sensitivity at all to the rate of reaction 7, since in this case other
reactions dominate the destruction of H$^{-}$.

\begin{table*}
\caption{Sensitivity of $J_{\rm crit}$ to the rate coefficient adopted for each reaction in the reduced network \label{tab:react}}
\begin{tabular}{clcc}
\hline
No.\ & Reaction &  Sensitivity (T4) & Sensitivity (T5) \\
\hline
1 & ${\rm H_{2} + \gamma} \rightarrow {\rm H + H}$ & 1.66 & 9.52 \\
2 & ${\rm H_2 + H} \rightarrow {\rm H + H + H}$ & 16.0 & 9.21 \\
3 & ${\rm H^{-} + H} \rightarrow  {\rm H_{2} + e^{-}}$ & 6.31 & 6.59 \\ 
4 & ${\rm H_{2}^{+} + H} \rightarrow {\rm H_{2} + H^{+}}$ & 1.10 & 1.02 \\ 
5 & ${\rm H^{+} + e^{-}} \rightarrow {\rm H + \gamma}$ & 15.3 & 18.1 \\
6 & ${\rm H + e^{-}} \rightarrow {\rm H^{-} + \gamma}$ & 7.01 & 33.3 \\
7 & ${\rm H^{-} + \gamma} \rightarrow {\rm H + e^{-}}$ & 5.54 & 1.04 \\ 
8 & ${\rm H + H^{+}} \rightarrow {\rm H_{2}^{+} + \gamma}$ & 3.89 & 1.40 \\ 
9 & ${\rm H_{2}^{+} + \gamma} \rightarrow {\rm H^{+} + H}$ & 1.09 & 1.00 \\
10 & ${\rm H + H} \rightarrow {\rm H^{+} + e^{-} + H}$ & 1.34 & 1.49 \\
11 & ${\rm H^{-} + H} \rightarrow {\rm H + H + e^-}$ & 1.12 & 3.98 \\ 
12 & ${\rm H + e^{-}}  \rightarrow {\rm H^+ + e^- + e^-}$ & 1.76 & 1.67 \\ 
13 & ${\rm H_{2}^{+} + He}  \rightarrow {\rm HeH^+ + H} $ & 1.00 & 1.00 \\
14 & ${\rm H + He}  \rightarrow {\rm H^+ + e^- + He} $ & 1.10 & 1.14 \\
15 & ${\rm H_2 + H^+}  \rightarrow {\rm H_2^+ + H} $ & 1.00 & 1.00 \\ 
16 & ${\rm H_2 + He}  \rightarrow {\rm H + H + He} $ & 1.00 & 1.00 \\
17 & ${\rm HeH^{+} + H}  \rightarrow {\rm H_2^+ + He}$ & 1.00 & 1.00 \\ 
18 & ${\rm H + H + H}  \rightarrow {\rm H_2 + H } $ & 1.00 & 1.00 \\ 
19 & ${\rm H^{-} + He}  \rightarrow {\rm H + He + e^-} $ & 1.00 & 1.02 \\
20 & ${\rm H_{2}^{+} + H} \rightarrow {\rm H + H^{+} + H}$ & 1.00 & 1.00 \\ 
21 & ${\rm He + H^{+}}  \rightarrow {\rm HeH^{+} + \gamma}$ & 1.00 & 1.00 \\
22 & ${\rm H^{-} + {\rm H^{+}}} \rightarrow {\rm H + H}$ & 1.05 & 1.75 \\
23 & ${\rm H_{2}^{+} + e^{-}} \rightarrow {\rm H + H}$ & 1.01 & 1.02 \\
24 & ${\rm HeH^{+} + e^{-}} \rightarrow {\rm He + H}$ & 1.00 & 1.00 \\ 
25 & ${\rm H^{-} + H^{+}} \rightarrow {\rm H_{2}^{+} + e^{-}}$ & 1.00 & 1.00 \\
26 & ${\rm H^{-} + e^{-}} \rightarrow {\rm H + e^{-} + e^{-}}$ & 1.00 & 1.00 \\ 
\hline
\end{tabular}
\\ The sensitivity quantifies the change in $J_{\rm crit}$ that occurs when we change the value of  the rate coefficient
by a factor of ten.
\end{table*}

\subsection{Assessing uncertainties in the reaction rates}
In the previous section, we showed that $J_{\rm crit}$ is highly sensitive to the values of the rates of
only a small number of the reactions included in our reduced chemical model. In this subsection, we
discuss how well we know the rates of these reactions. We restrict our attention to those reactions which
have sensitivities greater than 1.25 for either the T4 or the T5 spectrum (or both). 

\subsubsection{Photodissociation of H$_{2}$ (reaction 1)}
The molecular data on which the calculation of the H$_{2}$ photodissociation rate depends is
known accurately \citep[see e.g.][]{ar89}, with an error of no more than a few percent. Therefore,
in principle it should be possible to compute the H$_{2}$ photodissociation rate produced by
e.g.\ a T4 or T5 spectrum to a similar level of accuracy. However, in practice it is likely that the
H$_{2}$ photodissociation rates used in studies of the direct collapse model are rather less 
accurate than this, owing to the approximations made when deriving them.

One potential source of error comes from the value adopted for the H$_{2}$ photodissociation
rate in optically thin gas (i.e.\ the rate in the absence of self-shielding). To compute this, it is
necessary to specify the level populations of our collection of H$_{2}$ molecules. Most studies
of H$_{2}$ photodissociation in primordial gas assume that the gas is at low density and hence
that the effects of rotational and vibrational excitation can be ignored. In that case, the only energy
levels of H$_{2}$ that are populated are the $J=0$ para-hydrogen ground state and the
$J=1$ ortho-hydrogen ground state. Some calculations of the H$_{2}$ photodissociation rate
further assume that only the $J=0$ state is populated \citep[e.g.][]{abel97}, but in practice, if
only the lowest rotational levels are populated, then the H$_{2}$ photodissociation rate is
insensitive to the ortho-to-para hydrogen ratio.

\begin{figure}
\includegraphics[width=3.2in]{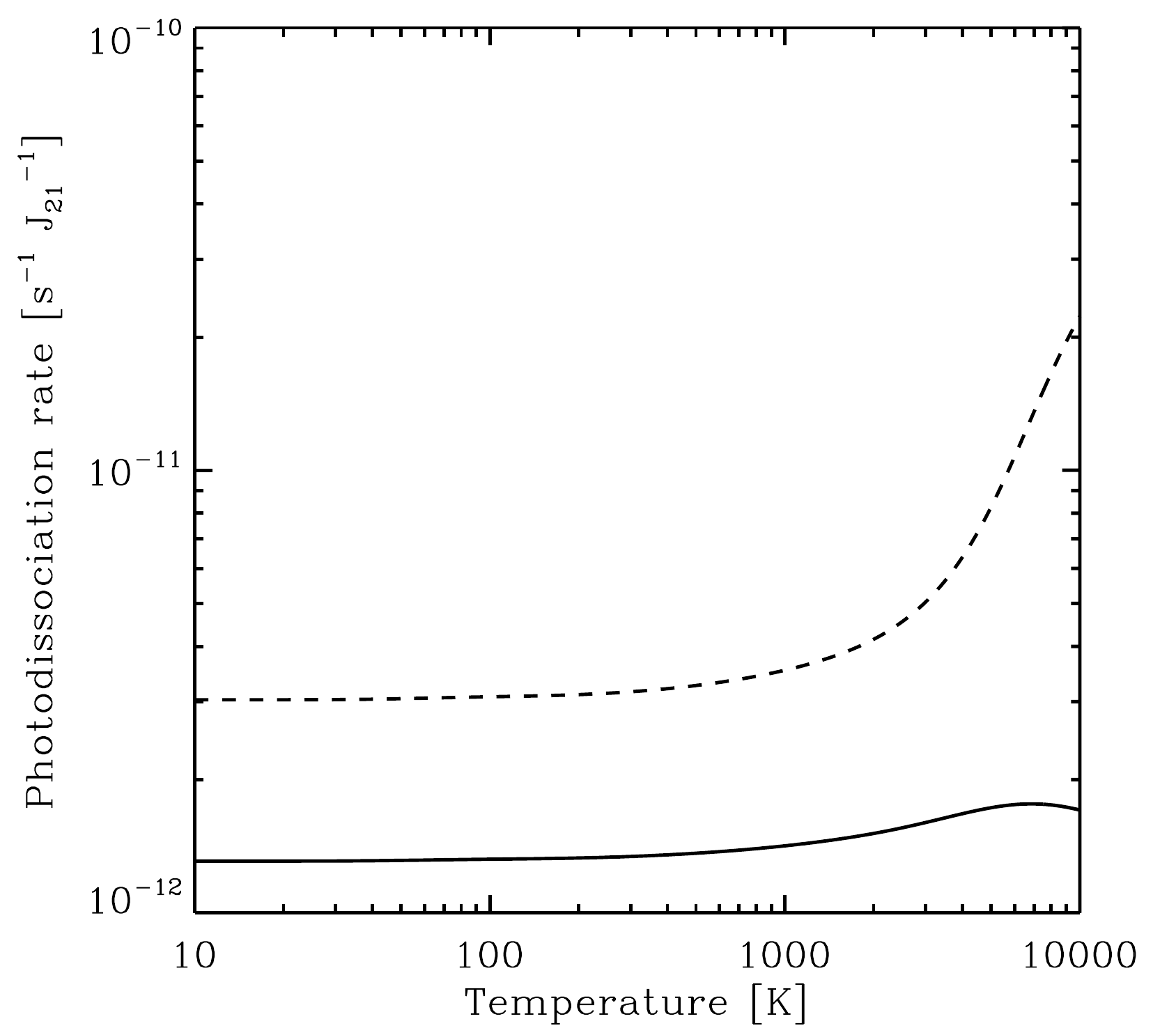}
\caption{Temperature dependence of the H$_2$ photodissociation rate in the LTE limit.
Values are shown for both a T4 spectrum (dashed line) and a T5 spectrum (solid line).
In both cases, we assume that $J_{21} = 1$.
\label{h2_pd_LTE}}
\end{figure}

In dense gas, however, it is no longer reasonable to assume that only the $J = 0$ and
$J = 1$ levels are populated. Instead, the populations of the excited rotational and vibrational 
levels will approach their LTE values. In LTE, the H$_{2}$ photodissociation rate increases
as the gas temperature increases, owing to the increasing importance of photodissociation
from levels with $v > 0$. This is illustrated in Figure~\ref{h2_pd_LTE}, where we plot the H$_{2}$ 
photodissociation rate in the LTE limit as a function of temperature, assuming either a T5
spectrum (solid line) or a T4 spectrum (dashed line). We see that in the case of a T5 spectrum,
the photodissociation rate of hot, thermally populated H$_{2}$ is around 30\% larger than the
rate at low temperatures, which in this case is the same as in the low density limit.
In the case of the T4 spectrum, the temperature dependence of the rate is much stronger: it 
increases by roughly a factor of seven as we increase the temperature from 10~K to $10^{4}$~K.

In order to help us assess how much influence the temperature dependence of the
optically thin H$_{2}$ photodissociation rate is likely to have on $J_{\rm crit}$, we have examined
the extreme case in which we adopt the LTE photodissociation rate throughout the collapse of the
gas. We find that in this case, $J_{\rm crit}$ is reduced to 12.3 if we use a T4 spectrum or
1210 if we adopt a T5 spectrum. Therefore, even in this extreme case, $J_{\rm crit}$ changes by
no more than around 30\%. In practice, the error introduced into $J_{\rm crit}$ by the use of
the low density limit for the optically thin H$_{2}$ photodissociation rate will probably be
smaller than this. This is because it is the behaviour of the gas at densities $n \sim 1000 \: {\rm cm^{-3}}$
that determines the value of $J_{\rm crit}$, and as this is less than the H$_{2}$ critical density, 
$n_{\rm crit} \sim 10^{4} \: {\rm cm^{-3}}$, the H$_{2}$ level populations will not yet have reached
their LTE values.
%these densities lie significantly below the
%H$_{2}$ critical density, $n_{\rm crit} \sim 10^{4} \: {\rm cm^{-3}}$, at which the H$_{2}$ level 
%populations approach their LTE values. 

 A second effect that can potentially become important when the photodissociation rate is extremely
 strong is the photodissociation of H$_{2}$ from highly excited vibrational levels. These
 levels can be populated by UV pumping of ground state H$_2$ \citep{shull78,db96} or
 by so-called formation pumping, i.e.\ the tendency for newly-formed H$_{2}$ to be in a
 highly excited rotational and vibrational state \citep[see e.g.][]{lau91}. The typical
 lifetimes of highly excited vibrational states of H$_{2}$ are short, of the order of 
 $10^{6} \: {\rm s}$ \citep{wol98}, and so photodissociation of H$_{2}$ from these states
 is significant only when the photodissociation rate is extremely large. This issue was
 investigated by \citet{shull78}, who showed that UV excitation of these highly excited
 vibrational levels becomes significant only for Lyman-Werner photon fluxes 
 $F > 3 \times 10^{-4} \: {\rm photons} \: {\rm cm^{-2} \: s^{-1} \: Hz^{-1}}$. If we
 convert this constraint to a constraint on $J_{21}$, we find that for a T5 spectrum
 it becomes approximately $J_{21} > 5 \times 10^{5}$, while for a T4 spectrum we have 
 $J_{21} > 10^{5}$. 
 
 These values of $J_{21}$ are much larger than the values of $J_{\rm crit}$
 derived using our one-zone model. However, one-zone calculations tend to underestimate
 $J_{\rm crit}$ compared to full 3D calculations, because they are unable to capture the
 complex temperature and density structure of the gas in real atomic cooling halos
 \citep[see e.g.\ the comparison of one-zone and 3D results in][]{sbh10}. 3D simulations
 performed using a T5 spectrum typically find that $J_{\rm crit} \sim \mbox{a few} \times
 10^{4}$ \citep{sbh10,latif15}, and models that account for the positive feedback from soft X-rays
 also find similarly high values \citep{it14}. At these values of $J_{21}$, radiative
 de-excitation of highly excited H$_{2}$ molecules remains more effective than UV excitation,
 but the latter effect could potentially contribute at around the 10\% level.  One effect of
 this will be to slightly suppress the H$_{2}$ formation rate, as some fraction of the newly
 formed, vibrationally hot H$_{2}$ molecules will be photodissociated before they can
 radiatively decay to the vibrational ground state. Another effect will be to slightly enhance
 the total H$_{2}$ photodissociation rate \citep{shull78}. A detailed accounting for the effect
 of this on the final value of $J_{\rm crit}$ is far outside the scope of the present work, but
 it seems likely that the effect will be of the order of 10\% or less. 
 
 Lastly, but most importantly, the values of $J_{\rm crit}$ that we derive in simulations
 performed using a T5 spectrum are highly sensitive to the treatment of H$_{2}$
 self-shielding adopted in the simulation \citep{whb11,soi14}. For ease of comparison
 to other recent calculations, the simulations presented in this paper were performed
 using the modified version of the \citet{db96} H$_{2}$ self-shielding function given in
 \citet{whb11}. This version of the self-shielding function is intended for use when one
 has rotationally hot H$_{2}$, i.e.\ when the H$_{2}$ rotational level populations are in
 LTE (although \citealt{whb11} assume that the vibrational level populations continue to
 be negligible). On the other hand, the original \citet{db96} self-shielding function, 
\begin{eqnarray}
f_{\rm sh, orig}(N_{\rm H_{2}}, T) & = & \frac{0.965}{(1 + x/b_{5})^{2.0}} + \frac{0.035}{(1+x)^{0.5}} \nonumber \\
& & \times \exp \left[-8.5 \times 10^{-4} (1+x)^{0.5} \right],
\end{eqnarray}
with $x = N_{\rm H_{2}} / 5 \times 10^{14} \: {\rm cm^{-2}}$ and $b_{5} = b / 10^{5} \: {\rm cm \: s^{-1}}$
gives a better fit when the H$_{2}$ is rotationally cold, with significant populations only in the $J=0$ and $J=1$ 
rotational levels. Depending on which version of $f_{\rm sh}$ one chooses, one recovers very
different values for $J_{\rm crit}$ in runs with a T5 spectrum \citep{soi14}. For example,
in our model, using $f_{\rm sh, orig}$ in place of $f_{\rm sh}$ increases $J_{\rm crit}$
from 1640 to approximately 9000.

This prompts the question of which version of the self-shielding function one should adopt
for these calculations. Unfortunately, this is not an easy question to answer. As we have already
discussed, at the density at which $J_{\rm crit}$ is determined in these simple one-zone calculations,
the H$_{2}$ level populations have not yet reached LTE. Nevertheless, the density is high enough
that collisional population of levels with $J > 1$ is starting to become important, and so we cannot
simply assume that the H$_{2}$ molecules are rotationally cold. It is likely that reality lies somewhere
between these two limiting cases, but how far in between is difficult to say. Clarifying this will
require one to  track the evolution of the H$_{2}$ rotational level populations during the collapse,
a task which lies outside of the scope of this present study.

Finally, in addition to this sensitivity to the choice of H$_{2}$ self-shielding function, the value
of $J_{\rm crit}$ is also sensitive to the method used to determine $N_{\rm H_{2}}$. Most one
zone calculations adopt the simple approximation $N_{\rm H_{2}} = n_{\rm H_{2}} L_{\rm char}$,
where $L_{\rm char}$ is some characteristic length scale, but different authors define this
different length scale in different ways. There is general agreement that it should be related
to the Jeans length, but no consensus on whether $L_{\rm char} = \lambda_{\rm J}$ or
$L_{\rm char} = \lambda_{\rm J} / 2$. Moreover, it is not always entirely clear what definition is
being used for $\lambda_{\rm J}$ -- different definitions in the literature can easily disagree
by a factor of two, depending on whether one derives $\lambda_{\rm J}$ from the dispersion
relation for a 1D plane wave or by considering the balance between thermal and gravitational
energy in a uniform density spherical cloud. Of course, in reality none of these prescriptions is
correct -- \citet{whb11} have shown that the Jeans length approximation can yield values of
$N_{\rm H_{2}}$ and $f_{\rm sh}$ that differ by an order of magnitude or more from the true
values that one derives by post-processing 3D simulations. To eliminate this source of error,
it is necessary to abandon the use of simple approximations for $N_{\rm H_{2}}$, and to compute
it self-consistently within a simulation using an algorithm such as {\sc TreeCol} \citep{cgk12}.
An attempt to do just this is reported on by \citet{hartwig15}, but their results lie
outside of the scope of the present study.

\subsubsection{Collisional dissociation of H$_{2}$ by H (reaction 2)} 
As discussed in e.g.\ \citet{msm96}, there are two separate processes that contribute to the rate of this reaction. First, there is direct collisional dissociation, involving
a transition from a bound state of the H$_{2}$ molecule directly into the continuum of classically unbound states. Second, there is dissociative tunneling, which involves
the excitation of  the H$_{2}$ molecule into a quasi-bound state -- a state that has an internal energy larger than is required for dissociation, but which is separated 
from the continuum by a barrier in the effective potential. H$_2$ molecules in these highly excited quasi-bound states dissociate only if they can tunnel through the
barrier before they have a chance to de-excite to a fully bound state. At temperatures $T < 4500$~K, this second process dominates the total collisional dissociation
rate \citep{mkm98}, and it remains important even at significantly higher temperatures. 
 Some previous studies of the direct collapse model have accounted only for the direct collisional dissociation of H$_2$, and not for the influence of dissociative tunneling
\citep[see e.g.][]{sbh10}. In paper I, we showed that the neglect of dissociative tunnelling can lead to one overestimating $J_{\rm crit}$ by up to a factor of two;
similar results have also been reported recently by \citet{latif14}.

Even if one does account for both processes, uncertainties in the rate coefficients adopted will also introduce some uncertainty
into $J_{\rm crit}$. In our simulations, we adopt rate coefficients for these processes based on the master equation study of
\citet{msm96}. They present complicated fitting functions for both rate coefficients that account for their dependence on both
the temperature and the density of the gas. The Martin et~al.\ study itself relies on state-to-state collisional rate coefficients
and collisional dissociation rates for H-H$_{2}$ collisions computed by \citet{mm93} using the quasi-classical trajectories (QCT) 
method and the LSTH potential energy surface \citep{l73,th78,sl78}. 

The use of the QCT method instead of a fully quantal method introduces some inaccuracy into the rate coefficients. At low temperatures, 
very large differences are found between the two methods \citep[see e.g.][]{sd94,bkmp2}, but much better agreement is found at high
temperatures, and so the QCT results should be reasonably reliable at the temperatures relevant for H$_{2}$ collisional dissociation.
Similarly, differences between the LSTH potential energy surface (PES) and more recent parameterizations of the H$_{3}$ PES 
\citep[e.g.][]{dmbe,bkmp1,bkmp2,mgp02} are important at low temperatures (compare, e.g., the results presented in
\citealt{sd94}, \citealt{lbd95} and \citealt{wf07}), particularly for pure rotational transitions, but appear to become much less important
at high temperatures \citep{mkm98}. In the absence of a comprehensive treatment of H$_{2}$ excitation by H along the lines of the
Martin et~al.\ study but using collisional rates computed with a different PES, it is difficult to state with certainty
the total error introduced into the collisional dissociation rate by the use of an older version of the PES. However, the comparison
between dissociation rate coefficients computed using the LSTH surface and the BKMP2 surface of \citet{bkmp2} shown in 
Figure~2 of \citet{ec09} suggests that the error must be relatively small at the temperatures of interest in this study. 
A conservative estimate for the error would be 25\%. This introduces an error into $J_{\rm crit}$ of roughly 35\% in simulations
with a T4 spectrum and 25\% in simulations with a T5 spectrum.
   
\subsubsection{Associative detachment of H$^{-}$ with H (reaction 3)}
The rate of this reaction has recently been measured  by \citet{kreck10} for
temperatures in the range $1 < T < 10^{4} \: {\rm K}$. The systematic error in
these measurements is approximately 25\%. The results agree well with the 
low temperature measurements of \citet{ger12} and the calculations of \citet{cizek98},
differing by amounts that are less than this systematic error. We have investigated
the effect of this uncertainty in the rate of reaction 3 by scaling the rate coefficient
both up and down by 25\% and determining what effect this has on our
estimate of $J_{\rm crit}$. In runs with a T4 spectrum, we find that 
$J_{\rm crit} = 15.1$ when we decrease the rate coefficient by 25\% and
$J_{\rm crit} =  21.6$ when we increase the rate coefficient by 25\%. In runs
with a T5 spectrum, the corresponding values are $J_{\rm crit} = 1350$ and 1930.
The remaining uncertainty in the rate of reaction 3 therefore introduces an
uncertainty of around 40\% into our estimates of $J_{\rm crit}$.

\subsubsection{Radiative recombination of H$^{+}$ (reaction 5)}
The rate of this reaction is known extremely accurately and high quality analytical fits are available that describe 
the temperature dependence of the rate coefficient. For example, the fit that we use in our study for the case B
recombination rate coefficient -- taken from \citet{hg97} and based on the data presented in \citet{fer92} -- has a quoted fitting error
of only 0.7\% over the temperature range $1 < T < 10^{9}$~K. Therefore, although $J_{\rm crit}$ is highly sensitive to the rate of this
reaction, the high accuracy with which the rate is known means that in general, this reaction does not contribute
significantly to the uncertainty in our estimates of $J_{\rm crit}$.

That said, one must still take care to use the {\em correct} recombination rate coefficient. Gravitationally
collapsing gas within an atomic cooling halo generally has a low fractional ionization, and hence the mean
free path for ionizing photons is small. For this reason, the on-the-spot approximation generally applies, and
when modelling the recombination of H$^{+}$ one should therefore use the case B recombination rate coefficient.
If the case A rate coefficient is used instead (as in e.g.\ the original \citealt{abel97} chemical model for primordial
gas, or the recent simulations by \citealt{latif15}), and the ionizing photons produced by direct recombination to the atomic hydrogen ground 
state are not explicitly accounted for, then the net effect is to increase the recombination rate by roughly 60\%
in the temperature range of interest.  To assess the effect that this would have on $J_{\rm crit}$,
we have re-run our simulations, using the case A recombination rate coefficient from \citet{hg97} in place of the
case B value. With this change, we find that $J_{\rm crit} = 10.1$ for a T4 spectrum and $J_{\rm crit} = 871$ for
a T5 spectrum, demonstrating that the use of the wrong rate coefficient introduces an 80--90\% error into 
our estimate of $J_{\rm crit}$.

\subsubsection{Radiative association of H and e$^{-}$ (reaction 6)}
\begin{figure}
\includegraphics[width=3.2in]{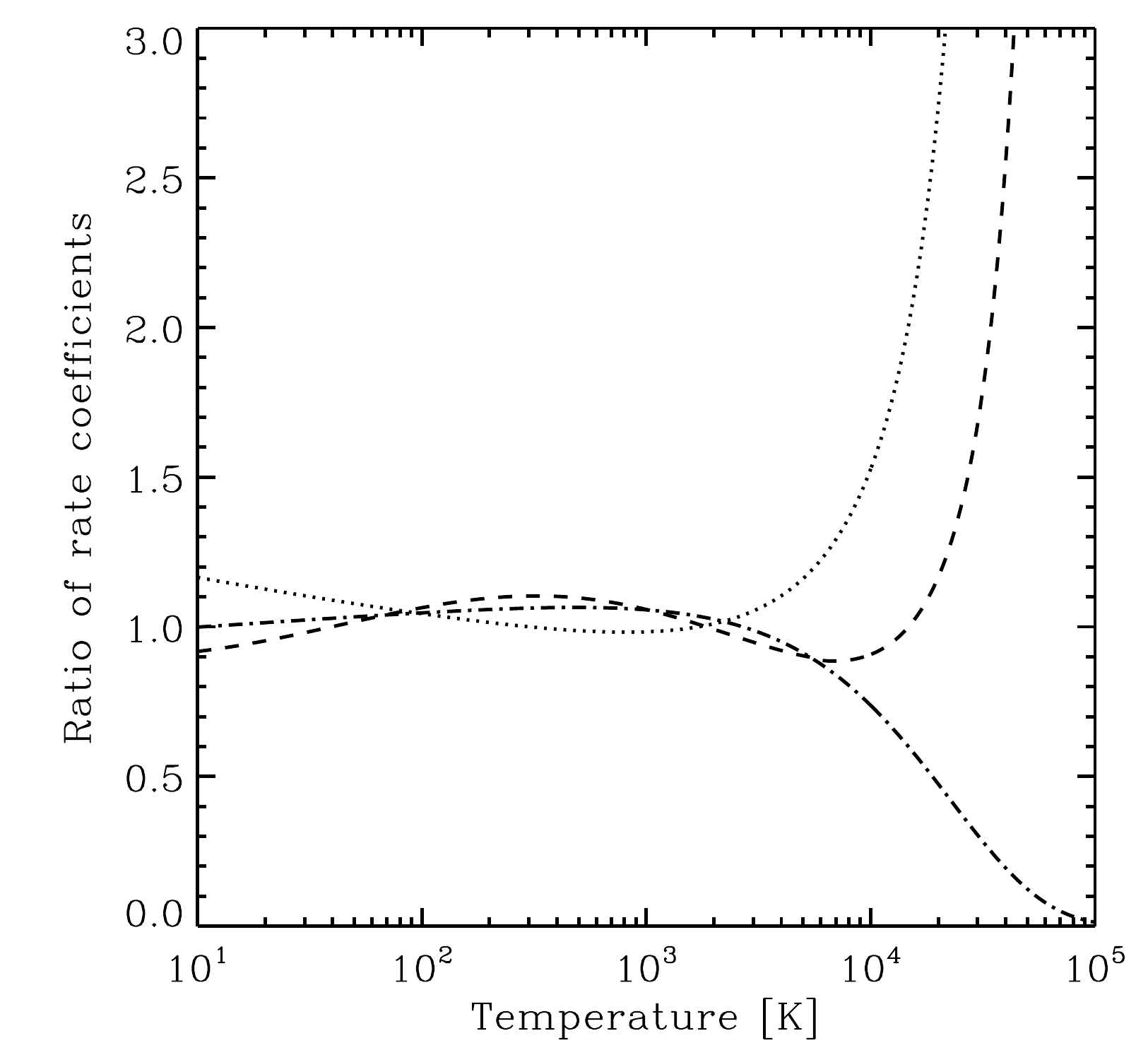}
\caption{Comparison of different determinations of the rate coefficient for the radiative association of H and e$^{-}$ 
(reaction 6). We show results for three different analytical fits, taken from \citet{hutch76} (dotted line), 
\citet{abel97} (dashed line), and \citet{sld98} (dot-dashed line). In each case, we plot the ratio of the rate coefficient given by
the analytic fits presented in these papers with that presented in the review of \citet{gp98}. \label{hmra}}
\end{figure}

Calculations of this rate in the literature generally derive it by applying detailed balance to the inverse process, H$^{-}$
photodissociation (reaction 7). Since the H$^{-}$ photodissociation cross-section is known accurately, one would expect
that there should be little uncertainty in the resulting rate coefficient. However, when constructing an analytical fit to the
calculated rate coefficient, it is necessary to choose some range of temperatures over which to fit. If the results of these
fits are then extrapolated outside of their range of validity, then there is no guarantee that the resulting rate coefficients
will remain accurate, and in this case, different fits may then yield very different results. This effect is seen quite clearly in
Figure~\ref{hmra}, where we compare several different analytical fits given in the literature. We plot results using the fits
given in \citet{hutch76} (dotted line), \citet{abel97} (dashed line) and \citet{sld98} (dot-dashed line), in each case showing the ratio of
the values given by these fits to those given by the fit in \citet{gp98}. 
%which we take as the default choice in our chemical model.

We see from Figure~\ref{hmra} that all four prescriptions agree well for gas temperatures in the range $10 < T < 2500$~K.
Within this temperature range, typical differences between different versions of the rate coefficient are of the order of
10\% or less. At temperatures $T > 2500$~K, however, the \citet{hutch76} fit starts to diverge strongly from the other rates.
This behaviour is unsurprising, as the range of validity quoted by \citeauthor{hutch76} for the fit is $100 < T < 2500$~K.

Similarly, at $T > 6000$~K, the other three versions of the rate coefficient start to differ significantly: the \citet{abel97}
version increases more strongly with increasing temperature than the \citet{gp98} version, while the \citet{sld98} version
of the rate coefficient increases less strongly than the \citet{gp98} version. This difference in behaviour is presumably
due to some or all of the analytical fits reaching the end of their temperature range of validity at $T \sim 6000$~K, 
although it is difficult to be certain as none of the other papers state the temperature range over which their fits are
supposed to be valid. 

We have investigated the effect that this difference in rates has on the value of $J_{\rm crit}$. For runs with a T4
spectrum, we find that $J_{\rm crit} = 16.9, 18.1, 20.0$ and 25.4 when we use the rate coefficient from \citet{sld98},
\citet{abel97}, \citet{gp98} or \citet{hutch76}, repsectively. For runs with a T5 spectrum, the corresponding values are 
$J_{\rm crit} = 1500, 1630, 1970$ and 2840.

We know from our own work in Paper I and from the work of previous authors \citep[e.g.][]{sbh10} that the value of $J_{\rm crit}$
is primarily determined by the behaviour of the H$_{2}$ fraction in gas with a density $n \sim 10^{3} \: {\rm cm^{-3}}$ and
a temperature $T \sim 7500$~K. In these conditions, the rate coefficient for reaction 6 given by \citet{hutch76} is not valid,
and consequently it is no surprise that the value of $J_{\rm crit}$ that we recover when we use this rate differs significantly
from those recovered using any of the other analytical fits. We do not recommend use of the \citet{hutch76} rate in future
studies of the direct collapse model.

As far as the other three fits are concerned, there seems to be no particular reason to prefer one over the others at the
present time. We can therefore conclude that the current uncertainty in this rate coefficient introduces an uncertainty of
between 20--30\% into the value of $J_{\rm crit}$.

Finally, it should be noted that the uncertainty in this rate coefficient may also affect the later thermal evolution of the
gas in halos where $J_{21} > J_{\rm crit}$, as in this case, H$^{-}$ bound-free emission becomes one of the most
important cooling processes once the density exceeds $n \sim 10^{8} \: {\rm cm^{-3}}$ \citep[see e.g.][]{om01,ssg10}.
However, a detailed study of this issue is outside of the scope of the present paper.
 
\subsubsection{Photodissociation of H$^{-}$ (reaction 7)}
Accurate cross-sections for this process are given by \citet{dejong72} and \citet{wish79}. 
More recently, \citet{miyake10} have computed revised cross-sections that account for
the effect of the prominent resonances in the cross-section at photon energies
$h\nu > 11$~eV. However, the effect of these resonances on the photodissociation rate
is small: for a T4 spectrum, the rate is enhanced by no more than a couple of percent,
while for a T5 spectrum, the enhancement is around 20\%. We have investigated the
effect of this enhancement on $J_{\rm crit}$, and find that for the T4 spectrum, it is
reduced by a few percent, to $J_{\rm crit} = 17.7$, while for the T5 spectrum, it is not
significantly affected. It should also be noted that this small chemical uncertainty is
dwarfed by the much larger astrophysical uncertainty arising from the fact that 
both the T4 and T5 spectra are relatively crude approximations to the spectra of real 
high-redshift protogalaxies \citep{soi14,ak15,latif15,agar15}.

\subsubsection{Radiative association of H and H$^{+}$ (reaction 8)} 
Rate coefficients for this reaction have been computed by both \citet{rp76} and \citet{sbd93}, and agree to within 3\%. 
Unfortunately, there is not such a good level of agreement between the different fits to the \citet{rp76} results that
have been used in the literature. \citet{sk87} introduced a fit of the form
\begin{equation}
k_{\rm 8, SK} = 1.85 \times 10^{-23} T^{1.8} \: {\rm cm^{3} \, s^{-1}}
\end{equation}
for $T < 6700$~K and
\begin{equation}
k_{\rm 8, SK} = 5.81 \times 10^{-16} \left(\frac{T}{56200} \right)^{-0.6657 \log (T / 56200)} \: {\rm cm^{3} \, s^{-1}}
\end{equation}
for $T > 6700$~K 
that was later adopted by many other authors; for instance, it is this prescription that is currently used in the
{\sc enzo} primordial chemistry model, as used in e.g.\ \citet{sbh10}. However, in their review of primordial
chemistry, \citet{gp98} use a different fit, ostensibly to the same data
\begin{eqnarray}
k_{\rm 8, GP} & = & {\rm dex} \left[-19.38 - 1.523 \log T + 1.118 (\log T)^{2} \right. \nonumber \\
& & \left. - 0.1269 (\log T)^{3} \right]  \: {\rm cm^{3} \, s^{-1}}.
\end{eqnarray}
Finally, in a recent paper, \citet{latif15} give the fit
\begin{eqnarray}
k_{\rm 8, L} & = & {\rm dex} \left[-18.20 - 3.194 \log T + 1.786 (\log T)^{2} \right. \nonumber \\
& & \left. - 0.2072 (\log T)^{3} \right]  \: {\rm cm^{3} \, s^{-1}},
\end{eqnarray}
citing \citet{cop11} as the source.

\begin{figure}
\includegraphics[width=3.2in]{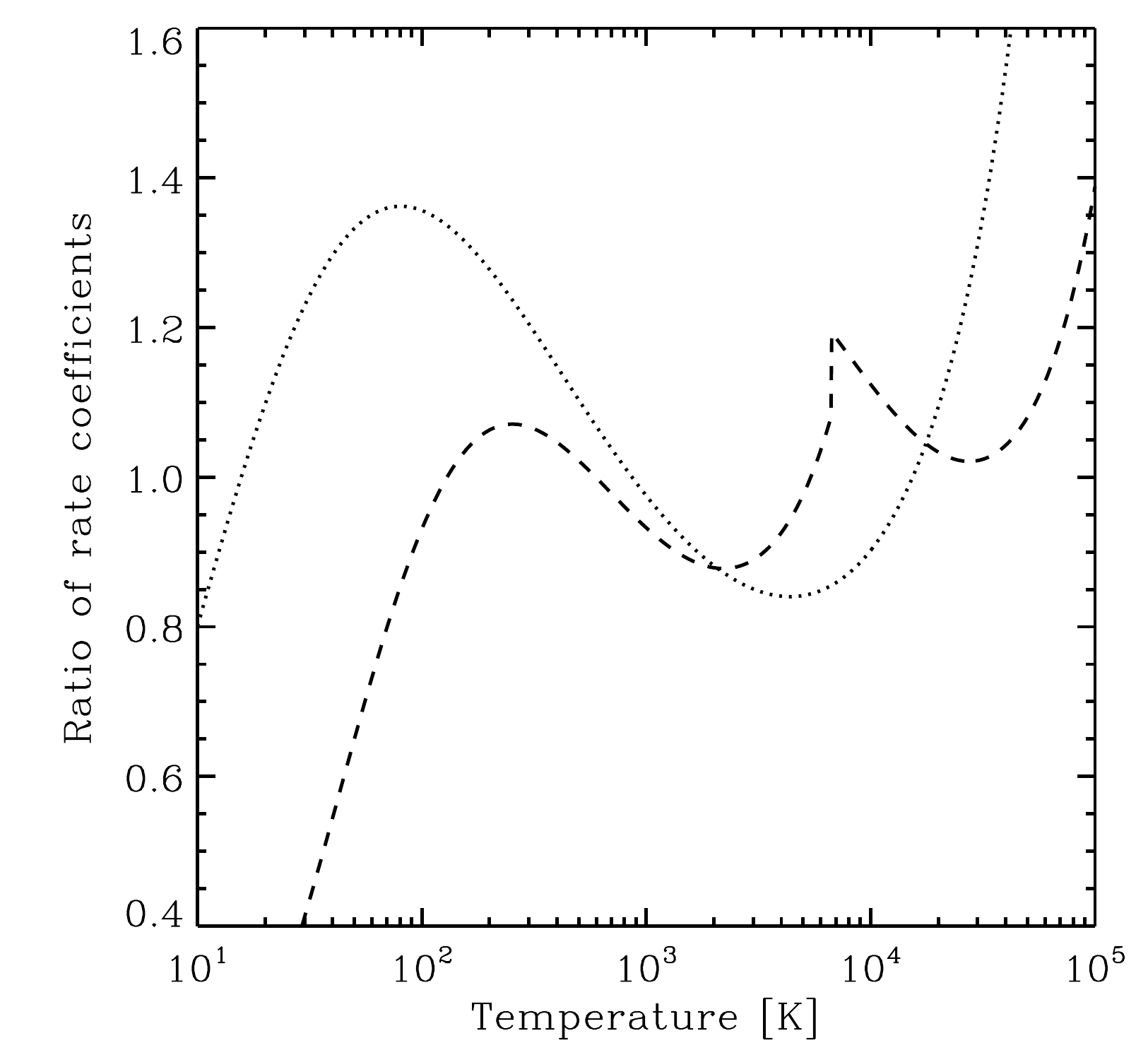}
\caption{Comparison of different analytical fits for the rate coefficient for the radiative association 
of H and H$^{+}$ (reaction 8). The dotted line shows the ratio of the fit given in \citet{gp98} to
that in \citet{latif15}, while the dashed line shows the ratio of the \citet{sk87} fit to the \citet{latif15}
fit. Comparison of the \citeauthor{latif15} fit with the rates tabulated in \citet{rp76} shows that it
agrees to within a few percent at the temperatures of interest. \label{h2pra}}
\end{figure}

In Figure~\ref{h2pra}, we compare these three different versions of the rate coefficient by plotting the ratios $k_{\rm 8, SK} / k_{\rm 8, L}$
and $k_{\rm 8, GP} / k_{\rm 8, L}$ as a function of temperature. We see that at the temperatures of interest, the scatter in the values
of the three fits is around 30\%, ten times larger than the difference between the \citet{rp76} and \citet{sbd93} calculations. This
scatter has an appreciable effect on the values of $J_{\rm crit}$ that we recover for runs performed using a T4 spectrum. We find
that for runs using the \citet{sk87}, \citet{gp98} and \citet{latif15} fits, $J_{\rm crit} = 19.8, 16.8$, and 18.0, respectively, a total 
difference of around 20\%. A similar analysis carried out for runs using a T5 spectrum finds a much smaller range of values:
for the same three cases, $J_{\rm crit} = 1670, 1600$, and 1630, giving us a total uncertainty in this case of only 4\%.

We have compared the values predicted by these three different versions of the rate coefficient with the values tabulated 
by \citet{rp76}, and find that the \citet{latif15} fit gives the best overall match to the tabulated data, agreeing to within around
4-5\% over the entire temperature range considered by \citeauthor{rp76} ($10 < T < 32000$~K). We therefore recommend
that this fit to the rate coefficient should be used in future numerical studies of the direct collapse model.

\subsubsection{Collisional ionization of H by H (reaction 10)}
\label{hhion}
Although there have been a number of calculations and measurements of the cross-section for this process at high
collision energies \citep[see e.g.][]{mcc68,hgg79,sbf89,rb99}, only a few studies have looked at its behaviour at the low energies relevant for the
direct collapse model.  In our fiducial model, we use the rate coefficient for this reaction given in \citet{lcs91}:
\begin{equation}
k_{\rm 10, LCS} = 1.2 \times 10^{-17} T^{1.2} \exp \left(-\frac{157800}{T} \right) \: \: {\rm cm^{3} \: s^{-1}}.
\end{equation}
This is based on the experimental cross-sections of \citet{gvz87}. However, their measurements only 
exist for collision energies $E > 100$~eV. Therefore, in order to derive a rate coefficient that is valid
at temperatures $T \sim 10^{4}$~K, it is necessary to extrapolate the measured cross-sections down
to the ionization threshold energy of 13.6~eV. This extrapolation can potentially introduce a substantial
error into the resulting rate coefficient. Indeed, a good example of this is provided by the fact that
\citet{hm89}, using the same cross-section data, derive a rate coefficient
\begin{equation}
k_{\rm 10, HM89} = 1.7 \times 10^{-14} T^{0.5} \exp \left(-\frac{149000}{T} \right) \: \: {\rm cm^{3} \: s^{-1}}.
\end{equation}
Although based on the same experimental data, these two expressions differ by a factor of 5 at $T = 10^{4}$~K
and by an order of magnitude at $T = 7000$~K.

The H-H collisional ionization rate coefficient can also be derived using the semi-empirical method described 
in \citet{drawin68,drawin69}. Following the interpretation of \citeauthor{drawin68}'s results given
in \citet{soon92}, we can write the rate coefficient as
\begin{equation}
k_{\rm 10, D69} = 1.57 \times 10^{-14} T^{0.5}  \psi(x) \: \: {\rm cm^{3} \: s^{-1}},
\end{equation}
where 
\begin{equation}
\psi(x) = \left(1 + \frac{2}{x} \right) \left[\frac{1}{1 + (2 m_{\rm e} / m_{\rm H} x)^{2}} \right] \exp(-x),
\end{equation}
$m_{\rm e}$ is the electron mass, $m_{\rm H}$ is the mass of a hydrogen atom,  and $x = 157800 / T$. 

Yet another expression for the H-H collisional ionization rate coefficient can be found in \citet{hm79}.
They assume that it is a constant factor of $1.7 \times 10^{-4}$ smaller than the H-e$^{-}$ rate
(citing \citealt{drawin69} in support of this fact), and hence write the rate coefficient as
\begin{equation}
k_{\rm 10, HM79} = 9.8 \times 10^{-15} T^{0.5} \exp \left(-\frac{157800}{T} \right) \: \: {\rm cm^{3} \: s^{-1}}.
\end{equation}

Finally, analytical expressions for the H-H ionization cross-section and corresponding thermal rate coefficient
are given in \citet{soon92}. These result from an extension of the work of \citet{ks91}, and were calculated
using the classical impulse approximation \citep{gryz65}. Unfortunately, there are a couple of significant
problems with the expressions presented in \citeauthor{soon92}'s paper, as already noted by \citet{barklem07}.
Most importantly, \citeauthor{soon92} uses the wrong value for the ionization threshold
energy. In the centre-of-mass frame, this should be equal to the ionization potential of hydrogen, $I_{\rm H} = 13.6$~eV.
However, Soon adopts a threshold which is half of this value, for reasons which are unclear \citep[see the
lengthy discussion in][]{barklem07}. In the high energy regime probed by experiments, this error is unimportant,
but at the low temperatures that we are interested in, the choice of threshold has a huge effect on the
rate coefficient, as we demonstrate below. In addition, \citeauthor{barklem07} reports that the rate coefficient
that one obtains by numerically integrating the ionization cross-section given by Soon (using Soon's choice of
threshold) differs significantly at $T < 10^{5}$~K from the collisional ionization rate coefficient given by Soon.

In view of these issues, we have chosen to proceed by ignoring the expressions given in \citet{soon92}
and instead analytically integrating the cross-section given in \citet{ks91}, using the correct value for the
energy threshold. At the low energies of interest here, this yields a rate 
\begin{equation}
k_{\rm 10, KS91} = 4.65 \times 10^{-21} T^{3/2} \exp \left(-\frac{157800}{T} \right) \: \: {\rm cm^{3} \: s^{-1}}.  \label{ksrate}
\end{equation}

\begin{figure}
\includegraphics[width=3.2in]{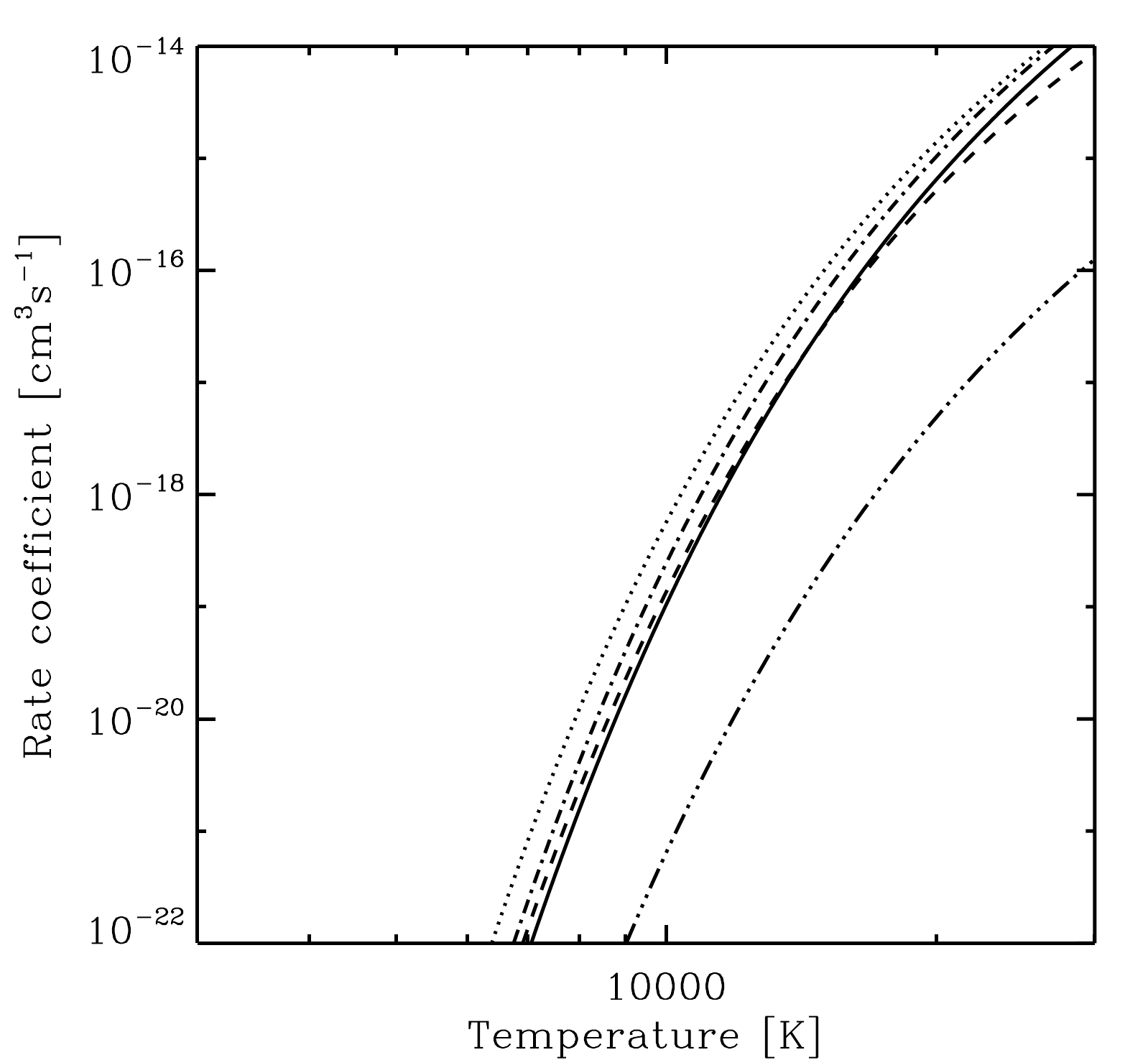}
\caption{Comparison of different rate coefficients for the collisional ionization of H by H (reaction 10).
The values shown are taken from \citet{lcs91} (solid line), \citet{hm89} (dotted line), \citet{hm79}
(lower dashed line), \citet{drawin69} (dash-dotted line), and this work, based on the cross-section
of \citet{ks91} (dot-dot-dot-dashed line).
\label{hhion_comp}}
\end{figure}

We compare these five different expressions for the H-H collisional ionization rate in Figure~\ref{hhion_comp}.
We see that the rate derived from the \citet{ks91} cross-section has an energy dependence that agrees
well with the other versions of the rate coefficient. However, there is a significant disagreement when it comes to the 
normalization of the rate, with this value being around a factor of 100 smaller than the others at all temperatures. 
Fortunately, the impact of this uncertainty on $J_{\rm crit}$ is relatively small.
%In this case, however, the impact on $J_{\rm crit}$ is not as pronounced. 
Using a small value for the rate coefficient of
reaction 10 has a similar effect to omitting it entirely, which as we saw in Paper I leads to a decrease in
$J_{\rm crit}$ of around 20-30\%. 

The remainder of the rate coefficients agree much better in terms of both their temperature dependence
and their normalization. Even so, there remains a substantial scatter in the values at low temperatures.
For example, as previously noted, the \citet{hm89} and \citet{lcs91} rates differ by around an order
of magnitude at $T = 7000$~K, despite being based on the same atomic data. This scatter introduces
a significant uncertainty into $J_{\rm crit}$: in runs with a T4 spectrum, the value of $J_{\rm crit}$ varies 
by up to 80\% depending on which version of the H-H collisional ionization rate coefficient we adopt,
while in runs with a T5 spectrum, the uncertainty is even larger, around a factor of 2.5

Finally, there remains the question of which of the different rate coefficients one should choose
for future calculations. As the \citet{ks91} cross-section agrees very well with the available experimental
data at $E > 100$~eV (see e.g.\ figure 3 in their paper), the rate coefficient based on this cross-section
is probably the most accurate one available at the present time. However, the fact that there is a large
gap between the range of energies for which the cross-section is experimentally constrained and the 
range of interest for deriving the low-temperature form of the rate coefficient means that we must treat
any of the possible rate coefficients with considerable caution.

\subsubsection{Collisional detachment of H$^{-}$ by H (reaction 11)}
Considerable uncertainty remains in the rate of this reaction. An early study by
\citet{bd69} gives two different sets of values for the reaction rate, depending on 
what form is chosen for the interaction potential. If the \citet{bard66a,bard66b}
potential is adopted,  then their numerical results are fit to within 10--20\% by 
the expression
\begin{equation}
k_{\rm 11, BD66} =  1.74 \times 10^{-12} T^{0.79} \exp \left(\frac{-8760}{T} \right) \: {\rm cm^{3} \: s^{-1}}.
\end{equation}
On the other hand, if the \citet{cp68} potential is used, we instead obtain a rate coefficient
that is well fit by the expression
\begin{equation}
k_{\rm 11, CP68} = 1.74 \times 10^{-13} T^{0.94} \exp \left(\frac{-8760}{T} \right) \: {\rm cm^{3} \: s^{-1}}.
\end{equation}
More recently, cross-sections for this reaction were given in the reviews of \citet{janev87} and
\citet{janev03}. Based on the \citet{janev87} data, \citet{lcs91} propose a rate coefficient of the
form
\begin{equation}
k_{\rm 11, LCS91} = 4.3 \times 10^{-16} T^{1.58} \: {\rm cm^{3} \: s^{-1}}.
\end{equation}
\citet{abel97} use the same cross-section data, but give a different analytical fit. At temperatures
$T_{\rm e} < 0.1$, where $T_{\rm e} = T / 11604.4 \: {\rm K}$, their fit has the form 
\begin{equation}
k_{\rm 11, A97} = 2.5634 \times 10^{-9} T_{\rm e}^{1.78186} \: {\rm cm^{3} \: s^{-1}},
\end{equation}
while at higher temperatures, it has the form
\begin{eqnarray}
k_{\rm 11, A97} & = & \exp \left[-20.37260896 \right. \nonumber \\
& & \mbox{} + 1.13944933 (\ln T_{\rm e}) \nonumber \\
& & \mbox{} -0.14210135 (\ln T_{\rm e})^{2} \nonumber \\
& & \mbox{} + 8.4644554 \times 10^{-3}  (\ln T_{\rm e})^{3} \nonumber \\
& & \mbox{}  - 1.4327641 \times 10^{-3}  (\ln T_{\rm e})^{4} \nonumber \\
& & \mbox{}   + 2.0122503 \times 10^{-4}  (\ln T_{\rm e})^{5} \nonumber \\
& & \mbox{}  + 8.6639632 \times 10^{-5}  (\ln T_{\rm e})^{6} \nonumber \\
& & \mbox{}  - 2.5850097 \times 10^{-5}  (\ln T_{\rm e})^{7} \nonumber \\
& & \mbox{}   + 2.4555012 \times 10^{-6}  (\ln T_{\rm e})^{8} \nonumber \\
& & \mbox{} \left. - 8.0683825 \times 10^{-8} (\ln T_{\rm e})^{9} \right] \: \: {\rm cm^{3} s^{-1}}.
\end{eqnarray}
Finally, the more recent data from \citet{janev03} can be used to derive a rate coefficient which is
fit to within 1\% over the temperature range $30 < T < 30000$~K  by the function
\begin{eqnarray}
k_{\rm 11, J03} & = & {\rm dex} \left[-10.911243 \right. \nonumber \\
& & \mbox{}  + 2.0648286 (\log T_{3}) \nonumber \\
& & \mbox{}  - 0.40506856 (\log T_{3})^{2} \nonumber \\
& & \mbox{} + 0.16462528 (\log T_{3})^{3} \nonumber \\
& & \mbox{}  - 0.10911689 (\log T_{3})^{4} \nonumber \\
& & \left. \mbox{}  +  0.022537886 (\log T_{3})^{5}\right] \: \: {\rm cm^{3} s^{-1}},
\end{eqnarray}
where $T_{3} = T / 1000$~K.

\begin{figure}
\includegraphics[width=3.2in]{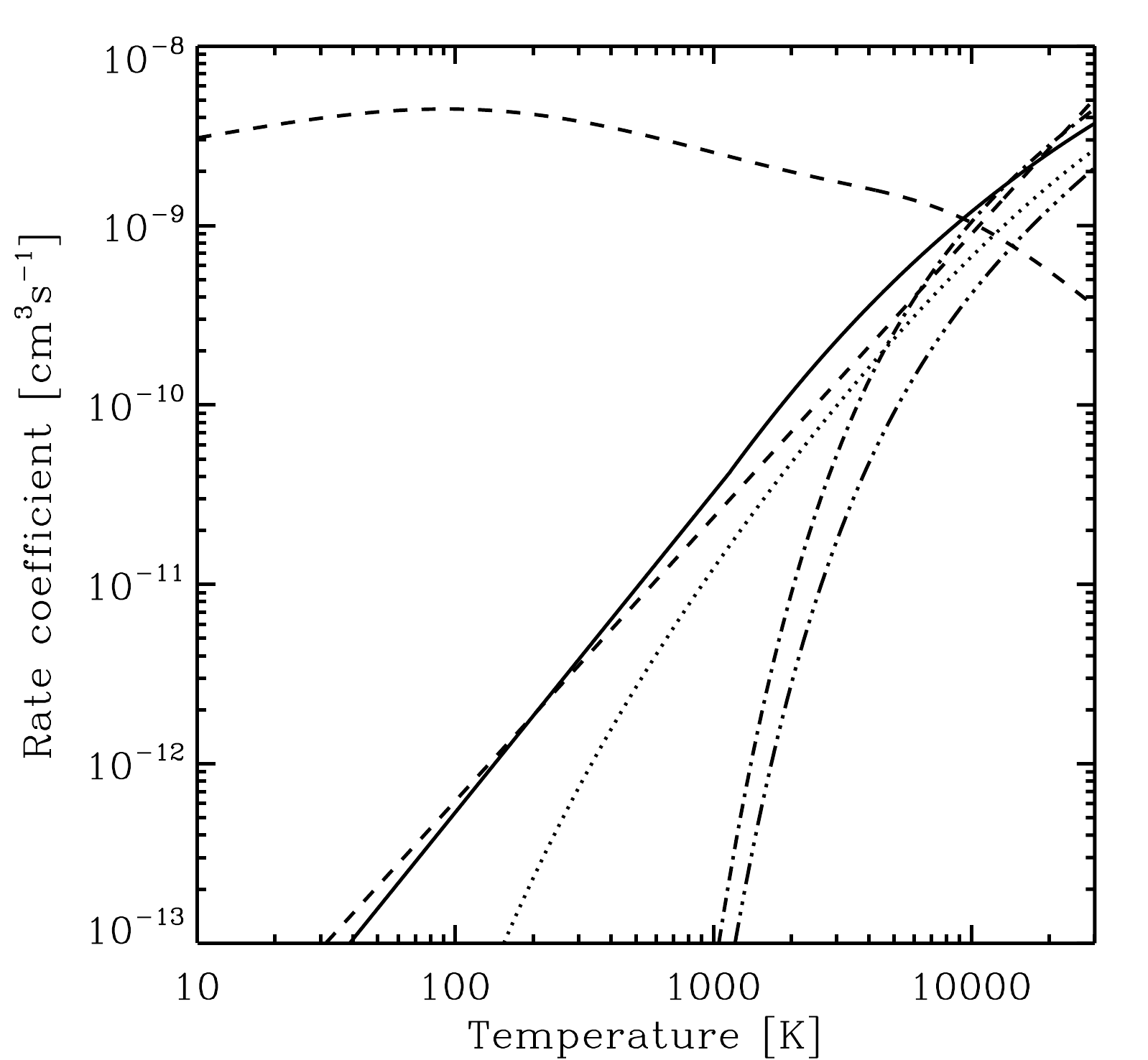}
\caption{Comparison of different rate coefficients for the collisional detachment of H$^{-}$ by
H (reaction 11). The values shown are taken from \citet{abel97} (solid line), our fit to \citet{janev03}
(dotted line), \citet{lcs91} (lower dashed line) and \citet{bd69}; in the latter case, the dash-dotted
line shows the results that they obtain for the  \citet{bard66a,bard66b} interaction potential,
and the dash-dot-dot-dotted line shows their results for the \citet{cp68} potential. For comparison,
we also plot the rate coefficient for reaction 3, the associative detachment of H$^{-}$ by H (upper
dashed line). The extremely large uncertainty in the rate of reaction 11 below a few thousand K is
unlikely to significantly affect the chemical evolution of the gas, as in this temperature regime, all of 
the different expressions yield values that are much smaller than the associative detachment rate.
\label{hmd}}
\end{figure}

In Figure~\ref{hmd}, we show how these different versions of the rate coefficient behave as a function
of temperature. We see immediately that at temperatures below a few thousand K, there is a large
uncertainty in the value of the rate coefficient. The values from \citet{bd69} fall off much more rapidly
with decreasing temperature than the other versions of the rate coefficient. Moreover, even if we 
ignore this early work, and restrict our attention to the results of more recent studies, we see that there
is still a large uncertainty in the low temperature behaviour, with the rate coefficient that we have derived
from the data in \citet{janev03} falling off more rapidly at low temperatures than the \citet{lcs91} or
\citet{abel97} rates. Fortunately, the chemical evolution of the gas is unlikely to be sensitive to this
large uncertainty. The reason for this is that at the temperatures where there is an order of magnitude
or more disagreement between the different rate coefficients, all of the expressions listed here predict
values that are much smaller than the associative detachment rate. Therefore, in this regime, associative
detachment dominates regardless of which expression we adopt for the rate of reaction 11.

Of greater concern is the smaller, but still substantial, disagreement between the values of the different
expressions at temperatures close to $10^{4}$~K. In this regime, the different determinations differ by
a factor of three to four, and the magnitude of the rate coefficient is similar to that for associative detachment,
meaning that this uncertainty potentially has a much larger impact on the H$^{-}$ abundance and the
H$_{2}$ formation rate. 

We have explored the effect that the uncertainty in this reaction has on the value of $J_{\rm crit}$. In
runs with a T4 spectrum, we find that the uncertainty introduced into our estimate of $J_{\rm crit}$ is
very small, of the order of a few percent. This is unsurprising, as in these runs, the dominant destruction
process for H$^{-}$ is photodissociation and so even large changes in the rate of reaction 11 have little
influence on the H$^{-}$ abundance and hence little impact on the H$_{2}$ formation rate. In runs
with a T5 spectrum, on the other hand, the uncertainty in the rate of reaction 11 has a much larger
effect, with $J_{\rm crit}$ varying between 1630 (if we use the rate from \citealt{abel97}) and 2600
(if we use the rate from \citealt{bd69} obtained with the \citeauthor{cp68} potential). The total uncertainty
in this case is therefore approximately 60\%. 

Although we would like to be able to recommend a particular choice of rate coefficient for this
reaction, it is highly unclear which of the different possibilities is likely to be the most accurate.
This process is therefore greatly in need of further theoretical or experimental study in order to
clarify the behaviour of its rate coefficient at temperatures $T < 10^{4}$~K.

\subsubsection{Collisional ionization of H by e$^{-}$ (reaction 12)}
Several different expressions for the rate coefficient of this reaction are in reasonably common
usage in the astrophysical literature. Previous studies of the direct collapse model have typically
adopted either the fit given in \citet{abel97} 
\begin{eqnarray}
k_{\rm 12, A97} & = & \exp \left[-32.71396786 \right. \nonumber \\
& & \mbox{} + 13.5365560 (\ln T_{\rm e}) \nonumber \\
& & \mbox{} - 5.73932875  (\ln T_{\rm e})^{2} \nonumber \\
& & \mbox{} + 1.56315498 (\ln T_{\rm e})^{3} \nonumber \\
& & \mbox{} - 0.28770560 (\ln T_{\rm e})^{4} \nonumber \\
& & \mbox{} + 3.48255977 \times 10^{-2} (\ln T_{\rm e})^{5} \nonumber \\
& & \mbox{} - 2.63197617  \times 10^{-3}  (\ln T_{\rm e})^{6} \nonumber \\
& & \mbox{} + 1.11954395  \times 10^{-4}  (\ln T_{\rm e})^{7} \nonumber \\
& & \mbox{} \left.  - 2.03914985 \times 10^{-6} (\ln T_{\rm e})^{8} \right] \: \: {\rm cm^{3} s^{-1}},
\end{eqnarray}
which is  based on data from \citet{janev87},  or the expression given in \citet{om01},
\begin{equation}
k_{\rm 12, O01} = 4.25 \times 10^{-11} T^{1/2} \exp \left(-\frac{157800}{T} \right) \: {\rm cm^{3} \, s^{-1}}
\end{equation}
which comes from \citet{lcs91}. On the other hand, \citet{hg97} give a different expression
\begin{equation}
k_{\rm 12, HG97} = \frac{21.11}{T^{3/2}} e^{-\lambda/2}
\frac{\lambda^{-1.089}}{(1 + [\lambda / 0.354]^{0.874})^{1.101}} 
\: \: {\rm cm^{3} \, s^{-1}}
\end{equation}
where $\lambda = 315614 / T$, which is their fit to \citet{lotz67}. Finally, \citet{sw91} give the expression
\begin{eqnarray}
k_{\rm 12, SW91} & = & \exp \left[-96.1443 \right. \nonumber \\
& & \mbox{} + 37.9523 \ln T \nonumber \\
& & \mbox{} - 7.96885 (\ln T)^{2} \nonumber \\
& & \mbox{} + 8.83922 \times 10^{-1} (\ln T)^{3} \nonumber \\
& & \mbox{} - 5.34513 \times 10^{-2} (\ln T)^{4} \nonumber \\
& & \mbox{} +1.66344 \times 10^{-3} (\ln T)^{5} \nonumber \\
& & \mbox{} -2.08888 \times 10^{-5} (\ln T)^{6} \nonumber \\
& & \mbox{} \left. - 157800 / T \right] \: \: {\rm cm^{3} s^{-1}},
\end{eqnarray}
which is based on the accurate cross-section measurements of \citet{seg87}. 

\begin{figure}
\includegraphics[width=3.2in]{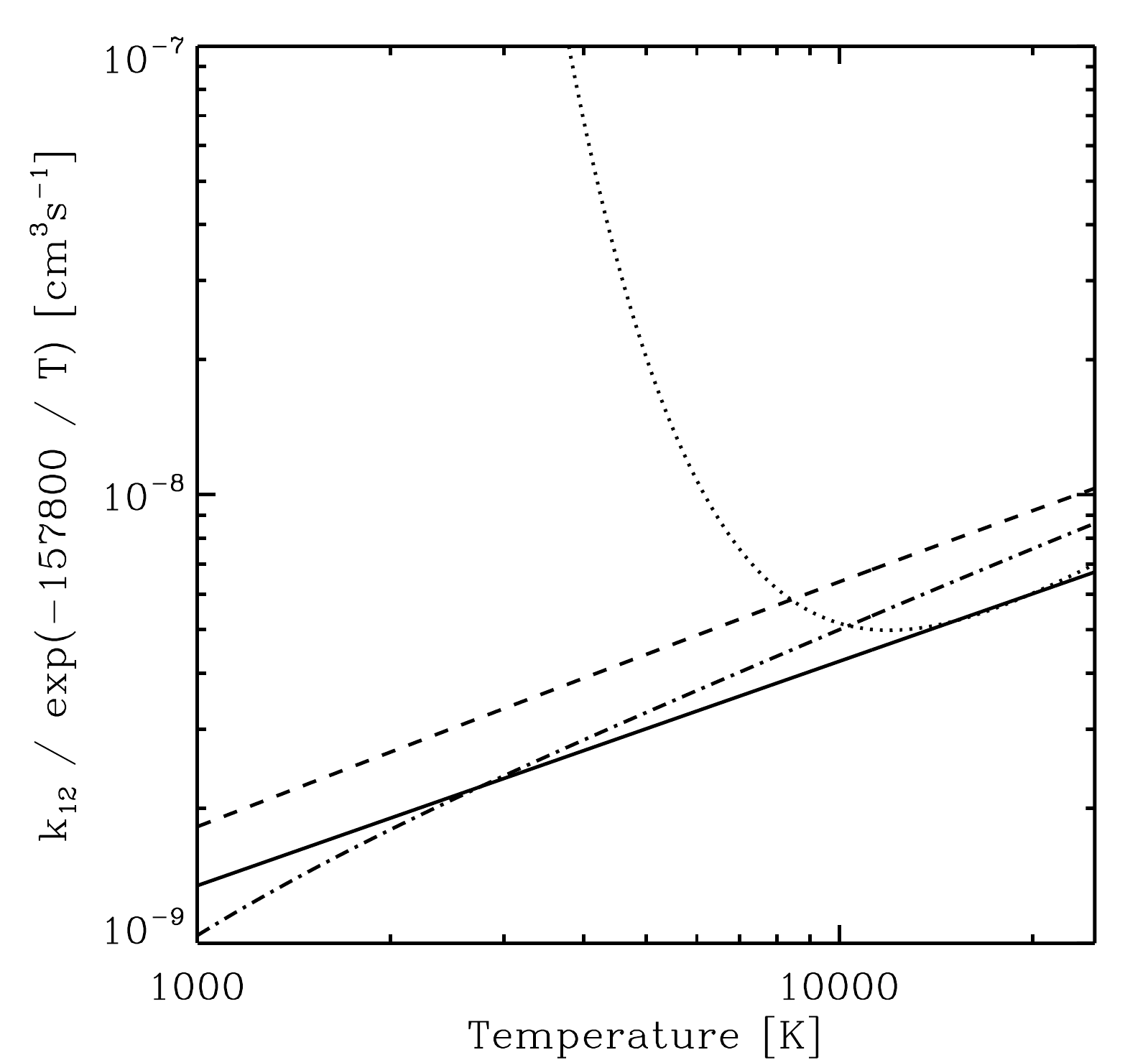}
\caption{Comparison of different rate coefficients for the collisional ionization of
H by electrons (reaction 12). To highlight the difference between the various
rate coefficients, we remove the underlying exponential temperature dependence
by plotting the quantity $k_{12} / e^{-157800/T}$, where $k_{12}$ is the rate
coefficient. The values shown in the plot come from \citet{lcs91} (solid line)
\citet{abel97} (dotted line), \citet{hg97} (dashed line) and \citet{sw91}
(dash-dotted line).
\label{hi_comp}}
\end{figure}

The behaviour of these different expressions as a function of temperature is compared in
Figure~\ref{hi_comp}. At low temperatures ($T \ll 10^{5} \: {\rm K}$), the collisional ionization rate has
an exponential dependence on temperature, $k_{12} \propto e^{-157800/T}$,  since only a 
small number of electrons in the exponential tail of the Maxwell-Boltzmann distribution have
sufficient energy to ionize hydrogen. Therefore, in order to allow us to better see the difference
between the different rate coefficients, we do not plot the rate coefficient $k_{12}$ in Figure~\ref{hi_comp},
but instead plot the value of $k_{12} / e^{-157800/T}$. From the Figure, we see that at temperatures 
$T > 10^{4} \: {\rm K}$, the different versions of the collisional ionization rate coefficient agree fairly
well. In this temperature regime, the different expression have roughly the same temperature 
dependence but disagree on the normalization of the rate coefficient by around 30\%. At lower
temperatures, the rates from \citet{lcs91}, \citet{sw91} and \citet{hg97} continue to agree fairly 
well, while the \citet{abel97} rate begins to disagree strongly with the others. It seems likely that
this is due to a problem with the \citeauthor{abel97} fit at these low temperatures, although it 
should be remembered that this occurs in a regime where the value of $k_{12}$ is falling off
exponentially with decreasing temperature, and so for most applications, the \citeauthor{abel97}
rate remains a suitable choice.

We have determined the value of $J_{\rm crit}$ that we obtain when using each of these
four expressions for the rate coefficient. In runs with a T4 spectrum, we find that the values
all lie in the range $J_{\rm crit} = 16.2$--18.0, while in the case of a T5 spectrum, the
corresponding range of values is $J_{\rm crit} = 1460$--1630. The uncertainty in the rate
of reaction 12 therefore introduces no more than a 10\% uncertainty into our estimates of
$J_{\rm crit}$.

Finally, which of these rate coefficients should one actually use in future calculations?
Since the total uncertainty (statistical and systematic) in the \citet{seg87} measurements
is unlikely to exceed 20\% \citep[see e.g.\ the discussion in][]{barklem11} and they are
also in good agreement with available high-accuracy theoretical calculations 
\citep[e.g.][]{bs93}, we recommend the use of the rate coefficient based on their 
measurements, i.e.\ the expression given in \citet{sw91}.

\subsubsection{Mutual neutralization of H$^{-}$ with H$^{+}$ (reaction 22)}
\begin{figure}
\includegraphics[width=3.2in]{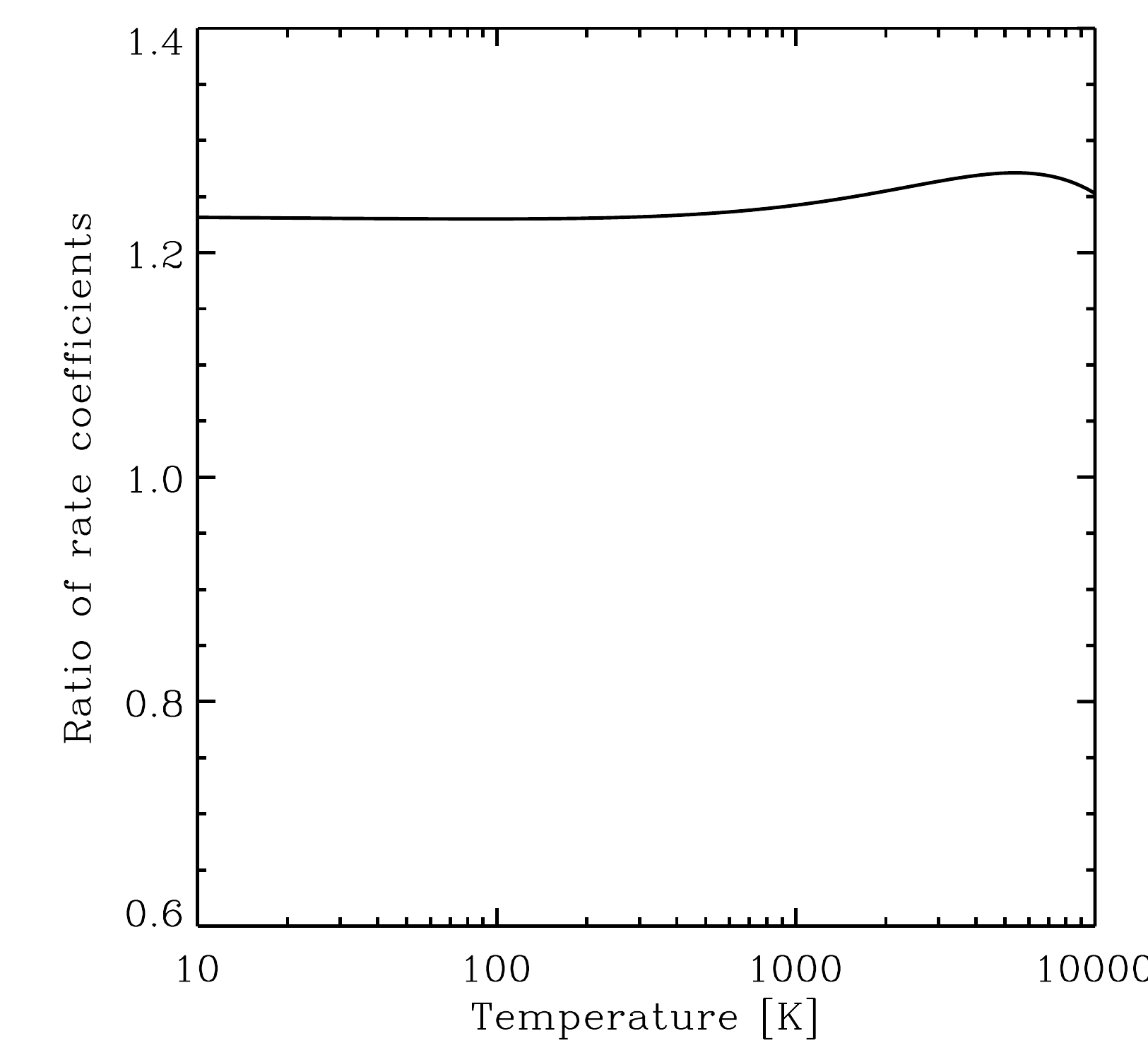}
\caption{Ratio of the rate coefficient for the mutual neutralization of H$^{-}$ by H$^{+}$
(reaction 22) given by \citet{sle09} to that given by \citet{cdg99}. 
\label{mn_comp}}
\end{figure}

In our fiducial model, we use the rate coefficient derived by \citet{cdg99}, based on the cross-sections of \citet{fk86}.
This is well fit by the simple function
\begin{equation}
k_{\rm 22, C99} = 2.4 \times 10^{-6} T^{-1/2}  \left(1 + \frac{T}{20000} \right) \: {\rm cm^{3} \: s^{-1}}.
\end{equation}
More recently, \citet{urbain12} state that their merged beam measurements ``support the rate coeffcient 
compiled by Croft et al.'', but do not give numerical details of their measured cross-sections, meaning that it is
not possible to assess how accurately their measurements agree with the earlier \citet{fk86} cross-sections.
The rate coefficient for reaction 22 has also been calculated relatively recently by \citet{sle09}. 
Their results are well fit in the temperature range $10 < T < 10000$~K by the function
\begin{eqnarray}
k_{\rm 22, SLE09} & = & 2.96 \times 10^{-6} T^{-1/2} - 1.73 \times 10^{-9}  \nonumber \\
& & \mbox{} + 2.50 \times 10^{-10} T^{1/2} \nonumber \\
& & \mbox{} - 7.77 \times 10^{-13} T  \: {\rm cm^{3} \: s^{-1}}.
\end{eqnarray}
We compare this with the Croft~et~al.\ rate in Figure~\ref{mn_comp}, where we plot the ratio 
$k_{\rm 22, SLE09} / k_{\rm 22, C99}$ for temperatures in the range $10 < T < 10000$~K. We see 
that the Stenrup~et~al.\ rate coefficient is higher than the Croft~et~al.\ rate coefficient at all temperatures, 
by around 20--25\%. We have investigated the effect of using the Stenrup~et~al.\ rate coefficient 
in place of the Croft~et~al.\ rate coefficient in our models. We find that if we do so, then there
is essentially no change in the value of $J_{\rm crit}$ we recover for runs performed using a T4 
spectrum. In runs with a T5 spectrum, however, we find that $J_{\rm crit} = 1630$ when we use the
Croft~et~al.\ rate coefficient and $J_{\rm crit} = 1550$ when we use the Stenrup~et~al.\ rate 
coefficient. The current uncertainty in the value of the mutual neutralization rate coefficient therefore
introduces an uncertainty of around 5\% into our derived value of $J_{\rm crit}$.

We note however that a number of treatments of primordial chemistry in the literature use expressions
for this rate coefficient based either on the measurements of \citet{map70} or taken from the review of
\citet{dl87}. These differ from the Croft~et~al.\ rate coefficient by factors of a few, as summarized in
\citet{gsj06}. We have explored the effect on $J_{\rm crit}$ of using one of these older values for
the rate coefficient, and find that if we use the \citeauthor{map70} value, we obtain 
$J_{\rm crit} = 17.4$ for a T4 spectrum and $J_{\rm crit} = 1070$ for a T5 spectrum, while if we
use the \citeauthor{dl87} value, we obtain instead $J_{\rm crit} = 18.3$  for a T4 spectrum and
$J_{\rm crit} = 1940$ for a T5 spectrum. We therefore see that the use of an older value for this
rate coefficient can introduce an error of up to $\sim 30$\% into the value we derive for $J_{\rm crit}$.

Because of this, and in view of the fact that the most recent determinations of the rate coefficient for reaction 
22 agree well with the Croft~et~al.\ value and disagree significantly with the \citet{map70} and \citet{dl87}
values, we recommend that the rate coefficients for this reaction from the latter two sources should no
longer be used when studying the direct collapse model.

\section{Impact of cooling rate  uncertainties on $J_{\rm crit}$}
\label{coolunc}
The cooling function that we use in our one-zone calculations accounts for the effects of radiative cooling from a large 
number of different coolants, including atomic hydrogen, atomic helium, H$_{2}$, H$_{2}^{+}$, and HeH$^{+}$
(see the detailed discussion in \citealt{Glover09}). However, in practice, the only coolants that contribute substantially
to the overall cooling rate are atomic hydrogen and H$_{2}$. We therefore focus our attention on these two coolants
and do not consider the effects of uncertainties in the other cooling rates, as these introduce a negligible uncertainty
into the total cooling rate.

\subsection{H$_{2}$ cooling}
There are two main sources of uncertainty that we need to consider in the case of H$_{2}$ cooling. 
At low densities, the cooling rate scales linearly with the collisional excitation rate, which is given
by the sum of a series of contributions corresponding to collisions between H$_{2}$ and H,
H$_2$ and He, H$_{2}$ and H$^{+}$, etc. In this regime, uncertainties in these contributions
directly affect the H$_{2}$ cooling rate. At high densities, on the other hand, the H$_{2}$ level 
populations reach LTE. In this regime, the H$_{2}$ cooling rate is insensitive to the collisional
excitation rate, and depends only on the partition function for the H$_{2}$ molecule, the energies
of the various excited rotational and vibrational levels, and their spontaneous radiative de-excitation
rates. All of these quantities are known accurately, and so in principle there should be almost no
uncertainty in the H$_{2}$ cooling rate in these conditions. In practice, however, the H$_{2}$
cooling rate in the LTE limit is often represented by a simple analytical function of temperature
that approximates the true H$_{2}$ cooling rate. Any error in this approximation therefore translates
into an uncertainty in the resulting parameterization of the H$_{2}$ cooling rate. 

\subsubsection{Low density limit}
In primordial gas, the most important collision partners responsible for exciting the internal
energy levels of the H$_{2}$ molecule are H and He atoms, other H$_{2}$ molecules, 
protons and electrons \citep{ga08}. In order to quantify which of these possible collisions
partners is most important in the present case, we have carried out a similar analysis to
that described in Section~\ref{howsense}. We constructed a series of different variants of
our cooling model by adjusting the individual collisional rate coefficients either upwards or
downwards by a factor of $10^{0.5}$ from their fiducial values.\footnote{The latter were  taken 
from \citet{ga08}, although the rates for H$_{2}$-H$^{+}$ and H$_{2}$-e$^-$ collisions were 
updated to account for new data, as summarized in Appendix A of Paper I.} We next used
our variant models to determine $J_{\rm crit}$ and then examined the difference in 
$J_{\rm crit}$ that resulted from adjusting each rate coefficient either upwards or downwards.
Finally, by taking the ratio of the largest and the smallest values of $J_{\rm crit}$ that we obtained
when modifying the rate of a particular collision process, we could define a sensitivity for that
process, analogous to the sensitivities determined for the different chemical rate coefficients
in Section~\ref{chemunc}.

\begin{table}
\caption{Sensitivity of $J_{\rm crit}$ to the rate coefficients adopted for the listed cooling processes \label{tab:cool}}
\begin{tabular}{lcc}
\hline
Process &  Sensitivity (T4) & Sensitivity (T5) \\
\hline
Lyman-$\alpha$ & 1.04 & 1.54 \\ 
H$_{2}$: collisions with H & 10.3 & 4.53 \\ 
H$_{2}$: collisions with He & 1.05 & 1.03 \\ 
H$_{2}$: collisions with H$_{2}$ & 1.00 & 1.00 \\ 
H$_{2}$: collisions with H$^{+}$ & 1.00 & 1.00 \\ 
H$_{2}$: collisions with e$^{-}$ & 1.00 & 1.00 \\
\hline
\end{tabular}
\end{table}

The results of our analysis are summarized in Table~\ref{tab:cool}. We see that $J_{\rm crit}$
displays a high sensitivity to the rate of only one of our considered collisional excitation 
processes, the excitation of H$_{2}$ by collisions with H atoms. This strongly suggests
that at the densities and temperatures where $J_{\rm crit}$ is determined, collisions between
H$_{2}$ and H dominate the total H$_{2}$ cooling rate. 

This behaviour differs from that
found in metal-free protogalaxies that are not illuminated by extremely strong radiation
fields, as in these systems H$_{2}$-He collisions play an important role, as do H$_{2}$-H$^{+}$
and H$_{2}$-e$^{-}$ collisions if the gas is cooling from an initially ionized state \citep{ga08}.
However, this difference is easy to understand. As we have already seen in paper I, in
atomic cooling haloes illuminated by a radiation field with $J_{21} \sim J_{\rm crit}$, the
critical factor that determines whether or not the gas can cool is the behaviour of the H$_{2}$
fraction at densities $n \sim 10^{3} \: {\rm cm^{-3}}$ and temperatures $T \sim 8000$~K.
In these conditions, the gas is mostly atomic and the fractional ionization is low, $x_{\rm H^{+}}
\sim 10^{-4}$. If we compare the mean H$_{2}$ cooling rates per collision due to collisions
with H, He, H$^{+}$ and e$^{-}$ at this temperature and density, then we find that 
$\Lambda_{\rm H_{2}, H} / \Lambda_{\rm H_{2}, He} \sim 4$ and 
$\Lambda_{\rm H_{2}, H} / \Lambda_{\rm H_{2}, H^{+}} \sim \Lambda_{\rm H_{2}, H} / \Lambda_{\rm H_{2}, e^{-}}
\sim 0.3$. Since the He:H ratio (by number) in primordial gas is approximately 0.08, it is clear
that collisions with helium contribute only a few percent of the total H$_{2}$ cooling rate,
and that the contributions from collisions with protons and electrons are negligible.
In the scenario considered by \citet{ga08}, gas cools and recombines from an initially ionized
state, but there is no ultraviolet background. In this case, H$_{2}$ cooling becomes effective
much earlier in the collapse, when the fractional ionization is significantly larger. In addition, 
\citet{ga08} were primarily concerned with the behaviour of the cooling rate at low temperatures,
where the difference between the contribution from H$_{2}$-H collisions and from collisions
with He, H$^{+}$ or electrons is much more pronounced. 

Since our analysis demonstrates that it is essentially just the H$_{2}$-H collisional excitation rate 
that plays a role in  determining $J_{\rm crit}$, we choose to focus our attention on this process, and 
will not discuss any uncertainties that may exist in the other collisional excitation rates. In our default
model, the treatment we use for the H$_{2}$-H contribution to the H$_{2}$ cooling rate is taken
from \citet{ga08}. They give the following analytic fit to the H$_{2}$ cooling rate, valid in the temperature
range $1000 < T < 6000$~K: 
\begin{eqnarray}
\Lambda_{\rm H_{2}, GA08} & = & {\rm dex} \left[ - 24.311209 \right. \nonumber \\
& & \mbox{} + 4.6450521 (\log T_3) \nonumber \\
& & \mbox{} - 3.7209846  (\log T_{3})^{2} \nonumber \\
& & \mbox{} + 5.9369081 (\log T_{3})^{3} \nonumber \\
& & \mbox{} - 5.5108047  (\log T_{3})^{4} \nonumber \\
& & \left. \mbox{} + 1.5538288 (\log T_{3})^{5} \right] \: \: {\rm erg \: s^{-1} \: cm^{3}},
\end{eqnarray}
where $T_{3} = T / 1000$~K.\footnote{\citet{ga08} also give fits for $\Lambda_{\rm H_{2}}$
in the temperature ranges $10 < T < 100$~K and $100 < T < 1000$~K, but the behaviour of
the H$_{2}$ cooling function at these temperatures has no influence on the value of $J_{\rm crit}$.}
This fit is based on the set of collisional excitation rate coefficients computed by 
\citeauthor{wf07}~(2007; see also \citealt{wgf07}). These authors use a fully quantal treatment
of the H$_{2}$-H system in their calculations and make use of the potential energy surface of
\citet{mgp02}. 

The other main treatment of H$_{2}$ cooling that is in widespread use in numerical
models of primordial gas is that of \citet{gp98}. They give the following analytical fit to 
the H$_{2}$-H cooling rate: 
\begin{eqnarray} 
\Lambda_{\rm H_{2}, GP98} & = &{\rm dex} \left[ -103.0 + 97.59 \log T - 48.05 (\log T)^2 \right. \nonumber \\
&  & \mbox{} - 48.05 (\log T)^2 + 10.80 (\log T)^{3}  \nonumber \\
& & \left. \mbox{} - 0.9032 (\log T)^{4} \right] \: \: {\rm erg \: s^{-1} \: cm^{3}}.
\end{eqnarray}
This fit is based on the quantal calculations of \citet{forrey97} at $T < 600$~K
and the quasi-classical calculations of \citet{mm93} at $T > 600$~K. In both cases, older
versions of the H$_{2}$-H potential energy surface were used, as summarized in \citet{gp98}.

Finally, an alternative version of the H$_{2}$-H cooling rate is given in \citet{msm96}. These
authors present a complicated fit to the H$_{2}$ cooling rate in the regime where H collisions
dominate that is valid over a wide range of different temperatures. However, the low density
limit of this fit is easy to extract and can be written as
\begin{eqnarray}
\Lambda_{\rm H_{2}, MSM96} & = &{\rm dex} \left[+195.09 - 137.9986  \log T \right. \nonumber \\
& & \mbox{} + 29.802 (\log T)^{2} \nonumber \\
& & \mbox{} -2.165163 (\log T)^{3} \nonumber \\
& & \mbox{} + 91.60106 \log (1.0 + 2680.075 / T) \nonumber \\
& & \mbox{} - 105.7438 \log (1.0 + 2889.762 / T)  \nonumber \\
& & \mbox{} \left. -4328.147 / T \right]  \: \: {\rm erg \: s^{-1} \: cm^{3}}.
\end{eqnarray}
This version of the cooling rate is also based on the calculations of \citet{mm93}.

\begin{figure}
\includegraphics[width=3.2in]{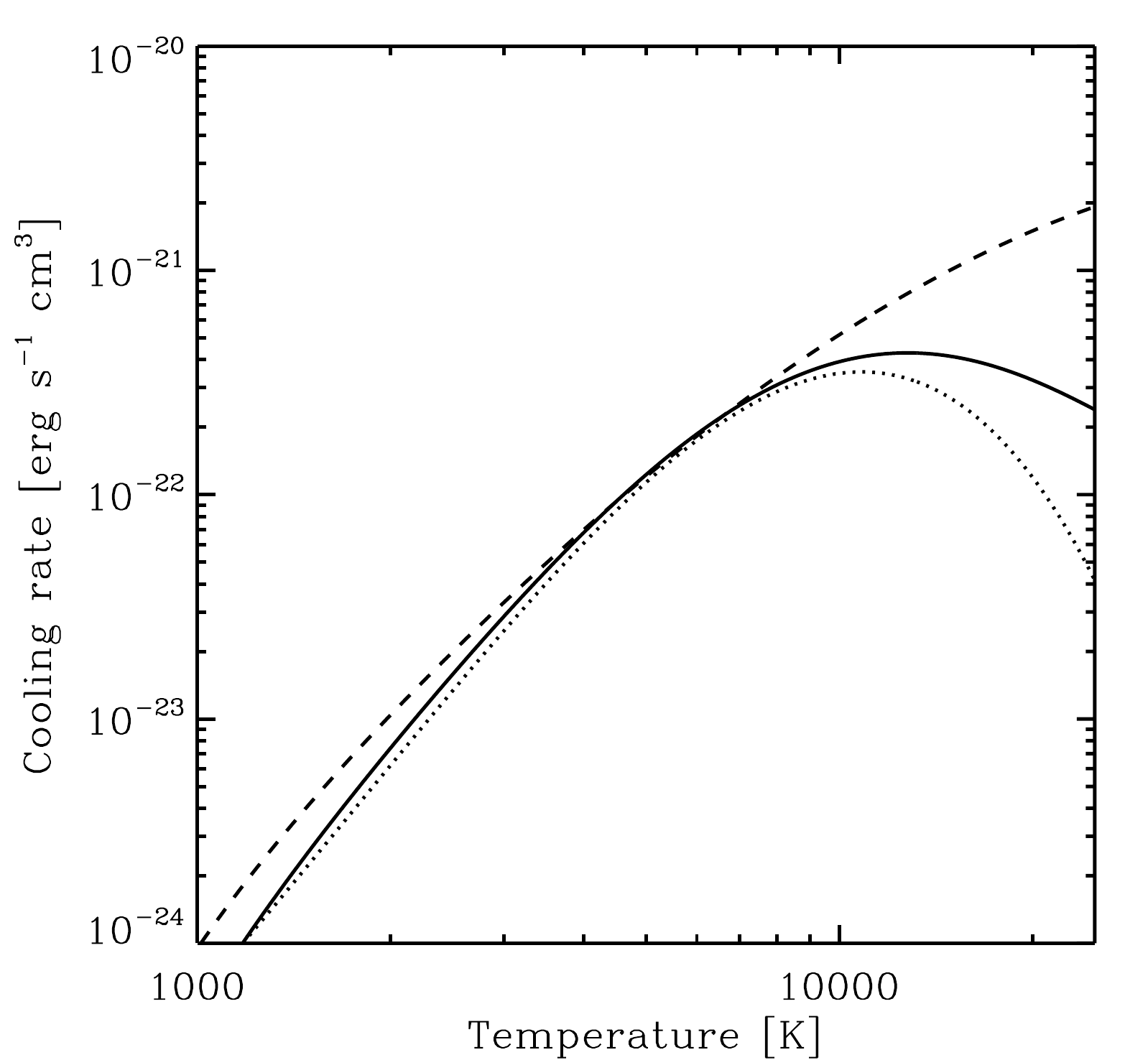}
\includegraphics[width=3.2in]{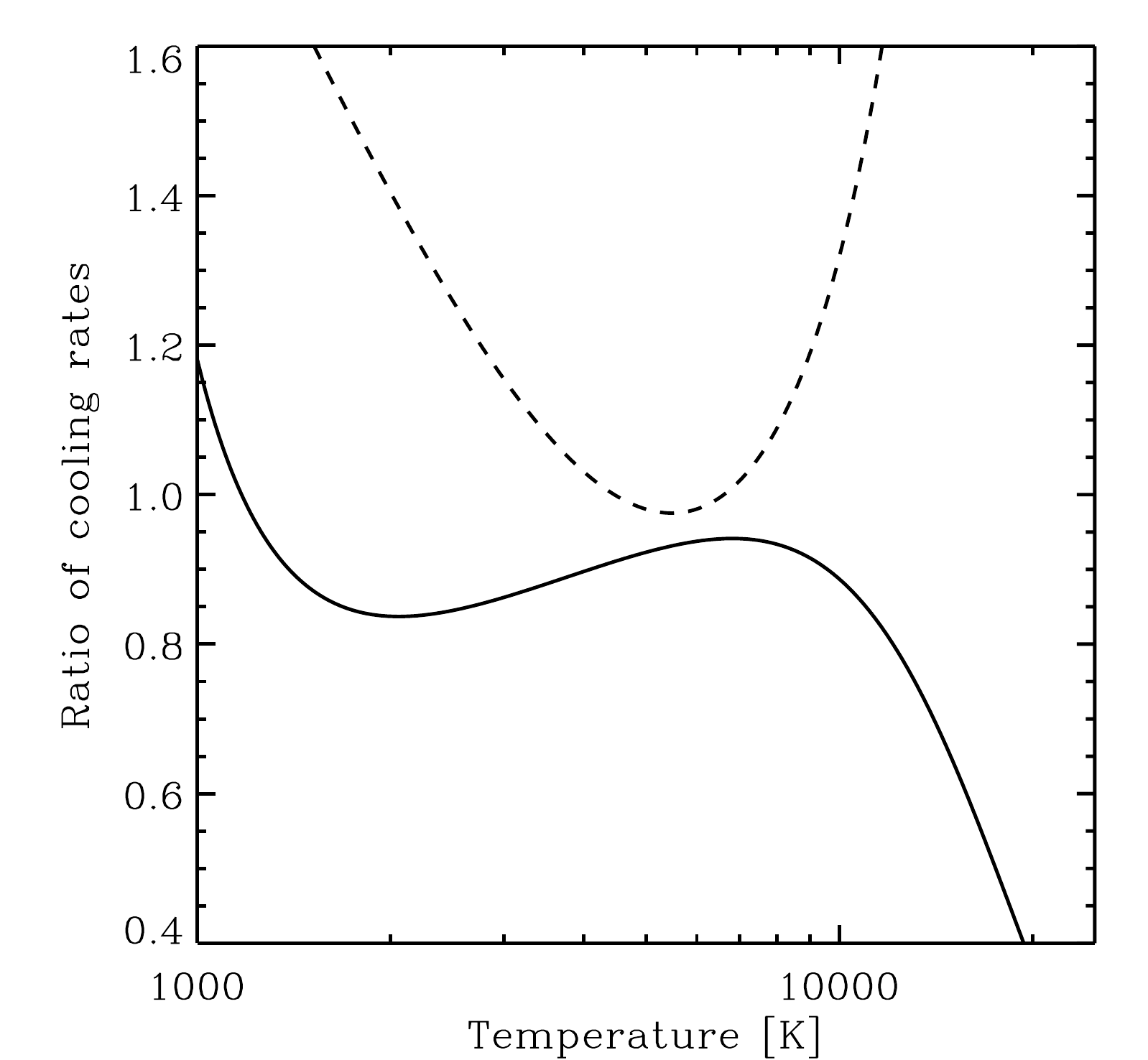}
\caption{{\it Upper panel}: comparison of H$_{2}$-H cooling rate coefficients.The solid line 
shows the high temperature fit from \citet{ga08}, the dotted line shows the expression from
\citet{gp98} and the dashed line shows the expression from \citet{msm96}. The significant
disagreement in behaviour at $T > 10^{4} \: {\rm K}$ is a consequence of the fact that the
fits given in \citet{ga08} and \citet{gp98} are not valid at these temperatures.
{\it Lower panel}: ratio of the \citet{gp98} rate to the \citet{ga08} rate (solid line) and
of the \citet{msm96} rate to the \citet{ga08} rate (dashed line).
\label{hot-h2-comp}}
\end{figure}

In the upper panel of Figure~\ref{hot-h2-comp}, we compare the temperature dependence of these 
three versions of the H$_{2}$-H cooling rate. We see that although the different rates disagree
somewhat at temperatures close to 1000~K, there is good agreement in the temperature range
$T \sim 6000$--8000~K relevant for $J_{\rm crit}$. At even higher temperatures, $T > 10^{4}$~K,
there is significant disagreement between the rates, but this is simply a consequence of the 
breakdown of the \citet{ga08} and \citet{gp98} analytical fits in this temperature regime; as
previously noted, the \citeauthor{ga08} fit is formally only valid at $T < 6000$~K, while the
\citeauthor{gp98} fit is valid only for $T < 10^{4}$~K. Fortunately, at these very high temperatures,
the precise behaviour of $\Lambda_{\rm H_{2}, H}$ is unlikely to be important, as Lyman-$\alpha$
cooling dominates in this temperature regime.

As the H$_{2}$ cooling rate varies over several orders of magnitude in the temperature range that
we are examining, it is difficult to tell from the upper panel of Figure~\ref{hot-h2-comp} exactly how
well the different cooling rates agree at $T \sim 6000$--8000~K. Therefore, in the lower panel of the
Figure, we plot instead the ratio of the \citeauthor{gp98} rate to the \citeauthor{ga08} rate (solid line)
and the ratio of the \citeauthor{msm96} rate to the \citeauthor{ga08} rate (dashed line). We see that
at the temperatures of interest, all three rates agree to within around 10\%.

We have investigated the impact of this small uncertainty on $J_{\rm crit}$. We find that in runs with
a T4 spectrum, it introduces an uncertainty of roughly 10\% into our determination of $J_{\rm crit}$,
while in runs with a T5 spectrum, the corresponding uncertainty is $\sim 5$\%.

\subsubsection{LTE limit}
In the LTE limit, the H$_{2}$ cooling rate can be computed very accurately, since it only depends
on the energies and radiative de-excitation rates of the rotational and vibrational levels of H$_{2}$,
all of which are known with a high degree of precision. \citet{glover11} numerically evaluated the LTE
cooling rate and produced the following analytical fit for the cooling rate coefficient, which is valid 
in the temperature range $100 < T < 10000$~K and has an error of no more than 5\% over this range:
\begin{eqnarray}
\Lambda_{\rm H_{2}, LTE, G11} & = & {\rm dex} \left[-20.584225 \right. \nonumber \\
& & \mbox{} + 5.0194035 \log T_{3} \nonumber \\
& & \mbox{} -1.5738805  (\log T_{3})^{2} \nonumber \\
& & \mbox{} - 4.7155769 (\log T_{3})^{3} \nonumber \\
& & \mbox{} + 2.4714161 (\log T_{3})^{4} \nonumber \\
& & \mbox{} + 5.4710750 (\log T_{3})^{5} \nonumber \\
& & \mbox{} - 3.9467356 (\log T_{3})^{6} \nonumber \\
& & \mbox{} - 2.2148338 (\log T_{3})^{7} \nonumber \\
& & \mbox{} \left. + 1.8161874 (\log T_{3})^{8} \right]  \: \: {\rm erg \: s^{-1}}.
\end{eqnarray}
However, many treatments of H$_{2}$ cooling in primordial gas \citep[see e.g.][]{gp98,bryan14,grassi14}
continue to make use of the approximate analytical expression for the H$_{2}$ LTE cooling rate given
in \citet{hm79}:
\begin{equation}
\Lambda_{\rm H_{2}, LTE, HM79} = \Lambda_{\rm H_{2}, rot} + \Lambda_{\rm H_{2}, vib}
\end{equation}
where
\begin{eqnarray}
\Lambda_{\rm H_{2}, rot} & = & \left(\frac{9.5 \times 10^{-22} T_{3}^{3.76}}{1 + 0.12 T_{3}^{2.1}} \right)
\exp \left[ - (0.13 / T_{3})^{3} \right]  \nonumber \\
& & \mbox{} + 3 \times 10^{-24} \exp(-0.51 / T_{3}) \: \: {\rm erg \: s^{-1}},
\end{eqnarray}
and
\begin{eqnarray}
\Lambda_{\rm H_{2}, vib} & = & \mbox{ } 6.7 \times 10^{-19} \exp (-5.86 / T_{3}) \nonumber \\
& & \mbox{} + 1.6 \times 10^{-18} \exp (-11.7 / T_{3}) \: \: {\rm erg \: s^{-1}}.
\end{eqnarray}

\begin{figure}
\includegraphics[width=3.2in]{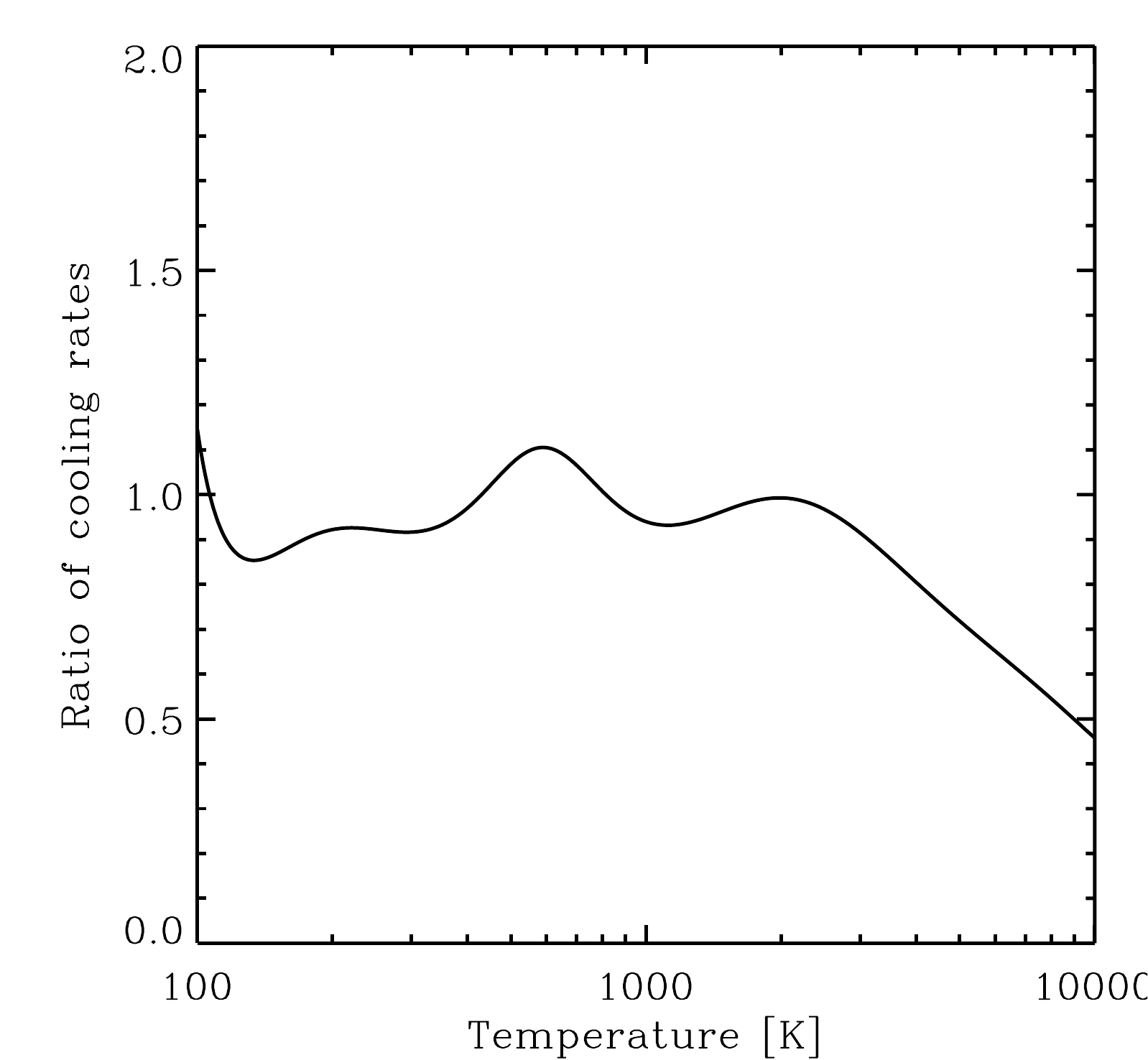}
\caption{Ratio of the accurate H$_{2}$ LTE cooling rate from \citet{glover11} to the approximate
analytical fit given in \citet{hm79}.  \label{H2_LTE_comp}}
\end{figure}

We compare these two cooling rates in Figure~\ref{H2_LTE_comp}, where we plot their ratio, 
$\Lambda_{\rm H_{2}, LTE, G11} / \Lambda_{\rm H_{2}, LTE, HM79}$.  We see that at temperatures
below $T \sim 2000$~K, the two rates agree reasonably well, with differences of no more than 10\%.
At higher temperatures, however, the expression from \citet{hm79} yields a larger cooling rate than the
one from \citet{glover11}, with the discrepancy worsening to as much as a factor of two as we near
$10^{4}$~K.

In practice, this discrepancy does not have a large impact on our estimate of $J_{\rm crit}$, as 
this is determined at densities somewhat lower than the H$_{2}$ critical density. If we compare the
results of runs performed using the \citet{glover11} LTE rate with those performed using the
\citet{hm79} rate, we find that $J_{\rm crit}$ differs by around 10\%. This holds regardless of whether
we use a T4 or a T5 spectrum.

\subsection{Lyman-$\alpha$ cooling}
\subsubsection{Collisions with electrons}
\begin{figure}
\includegraphics[width=3.2in]{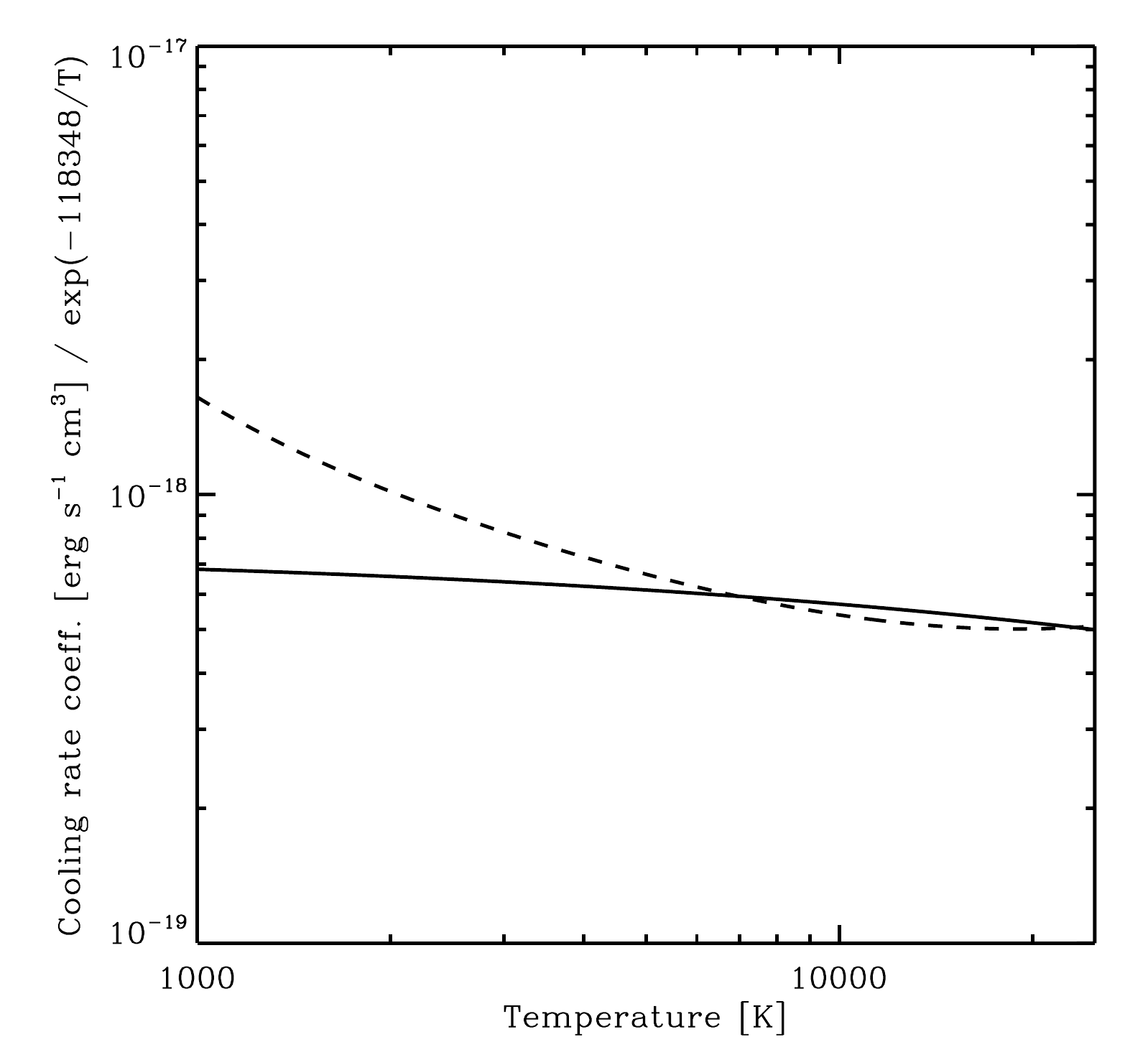}
\caption{Comparison of the Lyman-$\alpha$ cooling rate coefficients from \citet{cen92} (solid line)
and \citet{sw91} (dashed line). Both rate coefficients have been divided by a factor $e^{-118348/T}$ in order 
to remove the underlying exponential dependence on temperature, allowing us to more clearly  see
the difference in behaviour between the two expressions. \label{lyman_alpha_comp}}
\end{figure}

Most numerical models of primordial gas use the following expression for the cooling rate
coefficient arising from H-e$^{-}$ collisions (commonly referred to simply as Lyman-$\alpha$ cooling):
\begin{equation}
\Lambda_{\rm H, C92} = \left(\frac{7.5 \times 10^{-19}}{1 + T_{5}^{1/2}} \right) e^{-118348/T} \: \: {\rm erg \: s^{-1} \: cm^{3}}
\end{equation}
where $T_{5} = T / 10^{5}$~K. This expression comes from \citet{cen92}, and is a minor
modification of the cooling rate coefficient originally given in \citet{black81}. In order to
assess the level of uncertainty in this parameterization of the cooling rate, we compare it
with the expression given in \citet{sw91}:
\begin{equation}
\Lambda_{\rm H, SW91}  =  10^{-20} f(T) e^{-118348/T} \: \: {\rm erg \: s^{-1} \: cm^{3}},
\end{equation}
where
\begin{eqnarray}
f(T) & = & \exp \left[213.7913 - 113.9492 \ln T \right. \nonumber \\
& & \mbox{} + 25.06062 (\ln T)^{2} \nonumber \\
& & \mbox{} - 2.762755 (\ln T)^{3}  \nonumber \\
& & \mbox{} + 0.1515352 (\ln T)^{4}  \nonumber \\
& & \left. \mbox{}  - 3.290382 \times 10^{-3} (\ln T)^{5} \right].
\end{eqnarray}
These two versions of the Lyman-$\alpha$ cooling rate are compared in Figure~\ref{lyman_alpha_comp}.
To enable us to more clearly distinguish the differences in the behaviour of the two
rates, we have divided both of them by a factor $e^{-118348/T}$ in order to take
out the underlying exponential dependence on temperature. We see that although
there is some disagreement in the rates at low temperatures (where Lyman-$\alpha$
cooling is unimportant), at the temperatures relevant for $J_{\rm crit}$ these two
different prescriptions agree to within a few percent. This small difference introduces
an uncertainty of around 1\% into our estimate of $J_{\rm crit}$.

\subsubsection{Collisions with H and He atoms}
Electronic states of atomic hydrogen with $n > 2$ can also be excited by collisions with H and He
atoms. The rate coefficients for these processes are much smaller than for H-e$^{-}$ collisions,
but as in the case of the collisional ionization of hydrogen, it is conceivable that these processes
may become important if the fractional ionization of the gas is very small.

At the temperatures of interest in this study, the atomic hydrogen cooling rate is dominated by
the excitation of the $2p$ and $2s$ levels, and so there is no need for us to consider excitation
to states with $n \geq 3$. We can therefore write the cooling rate coefficient that represents the
effect of H-H  collisions as
\begin{equation}
\Lambda_{\rm H, H} = 10.2 \, {\rm eV} \times q_{12},   \: \: {\rm erg \, s^{-1} \, cm^{3}},
\end{equation}
where $q_{12}$ is the collisional excitation rate coefficient for transitions from $n = 1$ to 
$n = 2$.\footnote{In principle, one should distinguish between the $1s \rightarrow 2s$ and
$1s \rightarrow 2p$ transitions, but in practice, treatments of this process in the literature
often assume that both transitions share the same rate coefficient.}

As in the case of H-H collisional ionization, there are very few studies of H-H collisional
excitation that consider the low energies relevant for our present study. One possibility
is the rate given in \citet{lcs91}, which is based on the work of \citet{drawin69}:
\begin{eqnarray}
q_{12} & = & 5.8 \times 10^{-15} T^{1/2}  \left(1.0 + 1.69 \times 10^{-5} T \right) \nonumber \\
& & \times \exp \left(-\frac{118348}{T} \right) \: {\rm cm^{3} \: s^{-1}},
\end{eqnarray}
and so
\begin{eqnarray}
\Lambda_{\rm H, H, LCS91} & = & 9.5 \times10^{-26} T^{1/2}  \left(1.0 + 1.69 \times 10^{-5} T \right) \nonumber \\
& & \times \exp \left(-\frac{118348}{T} \right) \: \: {\rm erg \: s^{-1} \: cm^{3}}.
\end{eqnarray}

Collisional excitation of H by H is also treated by \citet{soon92}. However, in this case we encounter
the same problem as we did when examining the treatment of H-H collisional ionization by the same
author: the energy threshold adopted in Soon's calculations is only half of the value that
it should be \citep{barklem07}. If we 
correct for this problem and numerically integrate the cross-section given in \citet{soon92} using the
correct energy threshold, then
the cooling rate coefficient that we
obtain can be fit to within 10\% over the temperature range $1000 < T < 20000$~K by the function
\begin{equation}
\Lambda_{\rm H, H, S92b} = 5.8 \times 10^{-22} T^{-0.75} \exp \left(-\frac{118348}{T} \right) \: {\rm erg \: s^{-1} \: cm^{3}}
\end{equation}

\begin{figure}
\includegraphics[width=3.2in]{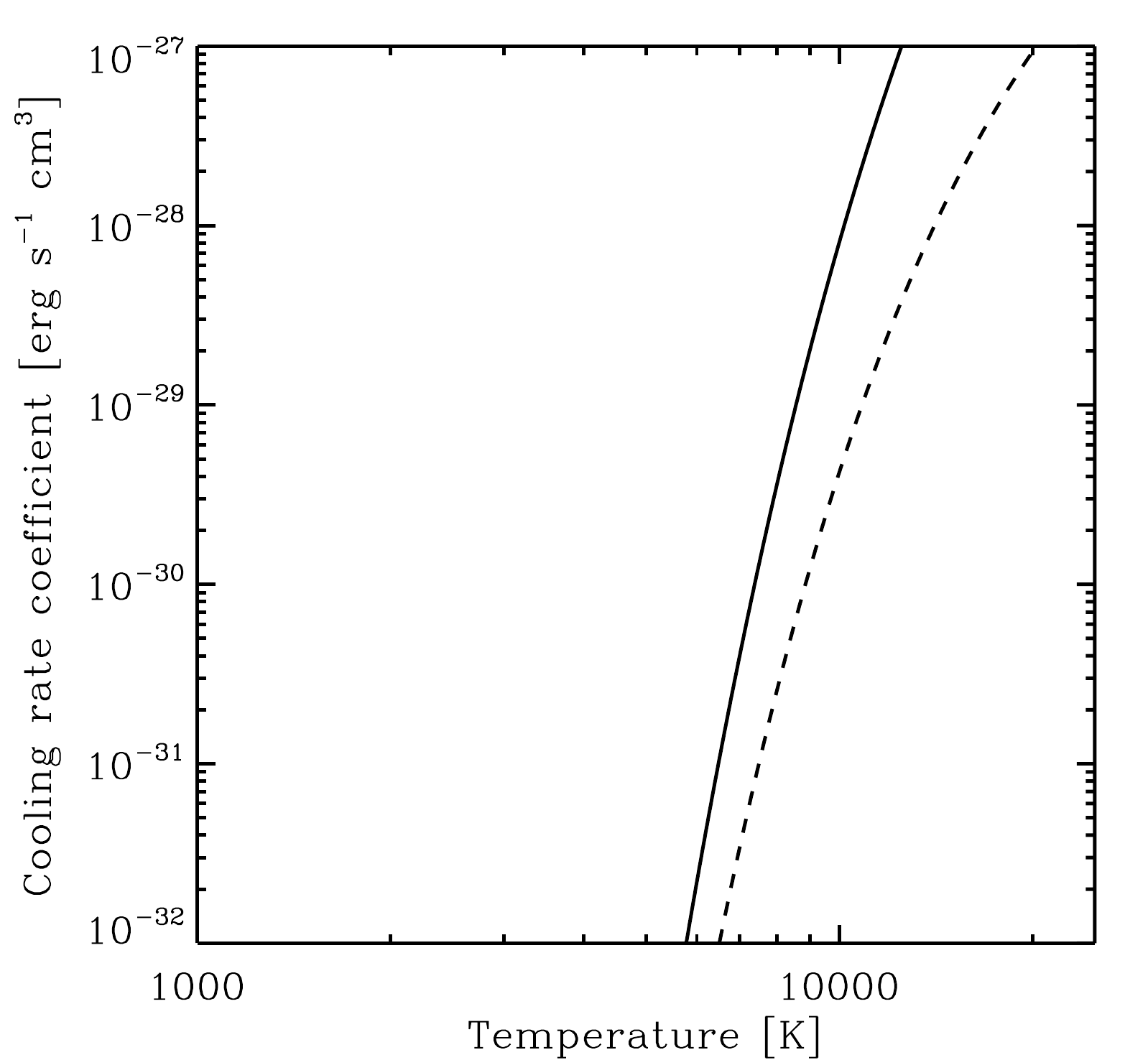}
\caption{Comparison of the H-H collisional excitation cooling rate coefficients from \citet{lcs91} (solid
line) and \citet{soon92} (dashed line).  \label{lyman_HH_comp}}
\end{figure}

In Figure~\ref{lyman_HH_comp}, we compare these two different H-H cooling rate coefficients. At the temperatures 
relevant for $J_{\rm crit}$, the rate coefficient from \citet{lcs91} predicts about an order of magnitude more cooling than the 
rate coefficient based on \citet{soon92}. 
If we compare these values to the H-e$^{-}$ cooling rate coefficients
discussed earlier, then we find that $\Lambda_{\rm H, H} / \Lambda_{\rm H, e^{-}} \sim 10^{-5}$ for the 
\citet{lcs91} rate coefficient and $\sim 10^{-6}$ for the rate coefficient based on \citet{soon92}. As the actual fractional ionization of the gas 
in the conditions where $J_{\rm crit}$
is determined is around $x \sim 4 \times 10^{-5}$ (see e.g.\ Paper I, Figure~1), it is unsurprising that if we include 
the effect of H-H cooling using either of the two rate coefficients from the literature, 
we find that the effect  on $J_{\rm crit}$ is negligible.
The large uncertainty in the H-H cooling rate therefore has no influence on the value of $J_{\rm crit}$.

As far as the excitation of H by collisions with He atoms is concerned, \citet{lcs91} give the following expressions
for the collisional excitation rates of the $1s \rightarrow 2p$ and $1s \rightarrow 2s$ transitions respectively:
\begin{eqnarray}
q_{\rm 1\rightarrow2p} & = & 1.2 \times 10^{-14} T^{1.2} \exp \left(-\frac{196250}{T} \right) \: \: {\rm cm^{3} \: s^{-1}}, \\
q_{\rm 1 \rightarrow2s} & = & 1.2 \times 10^{-15} T^{1.2} \exp \left(-\frac{196250}{T} \right) \: \: {\rm cm^{3} \: s^{-1}}.
\end{eqnarray}
These collisional excitation rates are based on the authors' fits to the data of \citet{gk84}. Using these excitation rates, 
we can easily derive the corresponding cooling rate coefficient:
\begin{equation}
\Lambda_{\rm H, He} = 2.2 \times 10^{-25} T^{1.2} \exp \left(-\frac{196250}{T} \right)  \: \: {\rm erg \: s^{-1} \: cm^{3}}.
\end{equation}
At $T \sim 10^{4}$~K, this is approximately $10^{5}$ times smaller than the expressions for $\Lambda_{\rm H}$
given above. Therefore, the cooling rate due to H-He collisions becomes comparable to that due to H-e$^{-}$
collisions only when the fractional ionization of the gas is very small, $x \sim 10^{-6}$. In the conditions
relevant for $J_{\rm crit}$, the fractional ionization is around a factor of 40 higher than this, and so it is
safe to conclude that H-He collisions make a negligible contribution to the total cooling rate.

\section{Discussion}
\label{discuss}
 
\subsection{Combining the uncertainties}
\label{comb}
It is useful at this point to draw a distinction between the uncertainties in $J_{\rm crit}$ that exist due
to the simplifications made in our chemical model and those that exist due to uncertainties in the 
chemical rate coefficient data. The case of H$_{2}$ self-shielding provides a good example of the former 
type of uncertainty. As we have already discussed, the value of $J_{\rm crit}$ that we recover in runs
with a T5 spectrum is highly sensitive to our choice of H$_{2}$ self-shielding function, increasing by
more than a factor of five if we use the original \citet{db96} self-shielding function instead of the more 
recent version given in \citet{whb11}. However, this large uncertainty ultimately stems from the fact that
these self-shielding functions are both approximations, valid in different limits. In reality, it is likely that
neither alternative provides a completely accurate description of the behaviour of the H$_2$ in the
gravitationally collapsing gas over the whole range of number densities of interest. Since the molecular
data needed to compute the H$_2$ photodissociation rate and self-shielding function is known accurately,
we can in principle eliminate this uncertainty simply by using a more sophisticated treatment of the H$_2$
in our simulation that tracks the population of the different rotational and vibrational levels and computes
the photodissociation rate separately for each level. Similarly, any uncertainties in the H$_{2}$ photodissociation
rate that are introduced by the simple approximations commonly used to compute $N_{\rm H_{2}}$ can be
eliminated by abandoning these approximations using a more accurate approach to compute the H$_{2}$
column density \citep[see e.g.][]{whb11,hartwig15}.

Uncertainties in $J_{\rm crit}$ that exist due to uncertainties in the chemical rate coefficient data are less
easily dealt with. In order to reduce their impact, it is necessary to determine the rate coefficients more
accurately, either experimentally or computationally, but this can be an extremely challenging and 
time-consuming occupation. It is therefore useful to assess the size of the uncertainty in $J_{\rm crit}$
that results from these rate coefficient uncertainties, so that we can better assess how important it is
to work on reducing these.  

\begin{table}
\caption{Error in $J_{\rm crit}$ introduced by uncertainty in the rate coefficients of the listed reactions
\label{tab:uncsum}}
\begin{tabular}{llcc}
\hline
& & \multicolumn{2}{c}{Uncertainty in $J_{\rm crit}$ (\%)} \\
No.\ & Reaction & T4 spec.\ & T5 spec.\  \\
\hline
2 & ${\rm H_{2} + H} \rightarrow {\rm H + H + H}$ & 35 & 25 \\
3 & ${\rm H^{-} + H} \rightarrow {\rm H_{2} + e^{-}}$ & 40 & 40 \\
6 & ${\rm H + e^{-}} \rightarrow {\rm H^{-} + \gamma}$ & 20 & 30 \\
10 & ${\rm H + H} \rightarrow {\rm H^{+} + e^{-} + H}$ & 80 & 120 \\
11 & ${\rm H^{-} + H} \rightarrow {\rm H + H + e^{-}}$ & Negligible  & 60 \\
\hline
\end{tabular}
%$^{*}$Note that we assume here that the unmodified version of the \citet{soon92}
%rate coefficient for this reaction is erroneous and hence neglect it, as discussed in Section~\ref{hhion}.
\end{table}

From the discussions in Sections~\ref{chemunc} and \ref{coolunc}, it is clear that in runs performed with a T4 spectrum, 
there are only four chemical reactions that introduce significant uncertainties into $J_{\rm crit}$: the collisional
dissociation of H$_{2}$ (reaction 2), the associative detachment of H$^{-}$ to form H$_{2}$ (reaction 3),
the formation of H$^{-}$ by radiative association (reaction 6) and the collisional ionization of H by H 
(reaction 10). These reactions individually introduce uncertainties of at least 25\% into $J_{\rm crit}$, as summarized 
in Table~\ref{tab:uncsum}, while the uncertainties contributed by the other processes discussed in Sections~\ref{chemunc} 
and \ref{coolunc} are at the level of 10\% or less. In runs with a T5 spectrum, the uncertainties introduced by these four reactions 
remain important, but $J_{\rm crit}$ also depends strongly on the rate of reaction 11, the collisional detachment of H$^{-}$ by 
collisions with H.

We can place some bounds on the total uncertainty introduced into $J_{\rm crit}$ by these reactions by deliberately 
choosing rate coefficients from amongst the different possibilities that either maximize or minimize $J_{\rm crit}$.
If we do this, then we find that in runs with a T4 spectrum, $J_{\rm crit, min} = 9.8$ and $J_{\rm crit, max} = 47.5$.
In runs with a T5 spectrum, we find instead that $J_{\rm crit, min} = 850$ and $J_{\rm crit, max} = 5540$.
Therefore, in the worst case, chemical rate coefficient uncertainties could introduce an uncertainty of around a factor
of five into our estimates of $J_{\rm crit}$.

In practice, the situation is probably not as bad as this, as the uncertainties in the different rate coefficients will cancel
with each other to some extent. To assess the importance of this, we recalculated $J_{\rm crit}$ a dozen times for both
the T4 and T5 cases, in each calculation using rate coefficients for reactions 2, 3, 6, 10 and 11 that were randomly selected 
from amongst the different possibilities.\footnote{In the case of reactions 2 and 3, we accounted for the 25\% systematic
uncertainty in the two rate coefficients by considering both our fiducial versions and also variants that were 1.25 and 0.75 
times these fiducial versions.} This procedure gave us 12 different values of $J_{\rm crit}$ for each choice of spectrum,
and we then calculated the mean and standard deviation of these sets of values. We found that for runs with a T4 spectrum,
$J_{\rm crit} = 21 \pm 5$, while for runs with a T5 spectrum, $J_{\rm crit} = 2400 \pm 700$. 
We can therefore conclude that the total uncertainty in $J_{\rm crit}$ due to uncertainties in the chemical rate coefficients
probably does not exceed a factor of two and certainly does not exceed a factor of five.

For comparison, determinations of $J_{\rm crit}$ made using 3D simulations rather than the simplified one-zone approach 
used here typically find that $J_{\rm crit}$ varies by at least a factor of a few from halo to halo owing to minor differences
in the details of the collapse of the gas and its resulting temperature structure \citep{sbh10,latif14}. Even larger
uncertainties are introduced if we abandon the use of our simplified T4 and T5 spectra and consider more realistic
models for the interstellar radiation field \citep{soi14,ak15,agar15}, particularly once we account for the fact that there may be
a non-negligible X-ray component \citep{it14,latif15}.
Therefore, it seems clear from the results of our study that astrophysical uncertainties have a larger impact on $J_{\rm crit}$
than chemical uncertainties.

\subsection{Comparison with previous work}
Previous studies of the direct collapse model using a one-zone approach have recovered a range of different values for
$J_{\rm crit}$, with values ranging from 18--39 in runs with a T4 spectrum, and 1400--16000 in runs with a T5 spectrum
(see the overviews of previous work in \citealt{it14} and \citealt{agar15}). Some of this uncertainty, particularly in the T5
runs, is due to differences in the treatment of H$_{2}$ self-shielding \citep{soi14}, and some can be ascribed to differences
in the composition of the chemical networks used to model the gas, as explored in Paper I. However, it is also useful to
assess how much of this uncertainty might reasonably be due to differences in the chemical rate coefficients used within
these different models. 

Specifically, in this section we compare the results that we obtain using our standard choices for the various rate coefficients
with those that we obtain using the set of rate coefficients adopted within (a) the {\sc enzo} primordial chemistry network
\citep{bryan14}, and (b) the {\sc krome} astrochemistry package \citep{grassi14}. We choose to compare our results with
these two alternative treatments because many of the previous studies of direct collapse use one or the other of them,
and moreover it is easy to gather information on exactly which rate coefficients are used within these models.

\begin{table}
\caption{Values of $J_{\rm crit}$ derived using the {\sc enzo} and {\sc krome} primordial chemistry networks \label{tab:compare}}
\begin{tabular}{lcc}
\hline 
&  \multicolumn{2}{c}{$J_{\rm crit}$} \\
Model & T4 spec.\ & T5 spec.\ \\ 
\hline
{\sc enzo} (unmodified) & 34.0 & 3050 \\   % Case A
{\sc enzo} (modified) & 24.1 & 2480 \\
{\sc krome} (unmodified) & 8.2 & 690 \\
{\sc krome} (modified) & 11.2 & 1080 \\  % Case A 
\hline
\end{tabular}
\end{table}

\subsubsection{The {\sc enzo} network}
Direct comparison between the results we obtain with our chemical network and those obtained using the {\sc enzo}
primordial chemistry network is complicated by the fact that the {\sc enzo} network not only makes different choices
for a number of the chemical and cooling rate coefficients, but also consists of a different set of chemical reactions
from those included in our model. In particular, the chemical network implemented in the most recent version of the 
{\sc enzo} code (version 2.4) omits two processes that have a significant influence on the value of $J_{\rm crit}$: the 
collisional ionization of H by H (reaction 10) and the dissociative tunneling contribution to the H$_{2}$ collisional 
dissociation rate. Therefore, to allow us to distinguish between the differences in $J_{\rm crit}$ caused by the omission
of these reactions and the differences in $J_{\rm crit}$ caused by the differences in the rate coefficients, we compare
our results with those from two additional sets of calculations. In one, we use the unmodified {\sc enzo} network. In
the other, we use a version of this network that we have modified to include the two omitted processes mentioned
above. 

The results of our comparison are shown in Table~\ref{tab:compare}. We see that the {\sc enzo} network systematically
produces higher values of $J_{\rm crit}$ that those derived using our network. With the unmodified {\sc enzo} network,
$J_{\rm crit}$ is increased by almost a factor of two for both the T4 and the T5 spectrum. With the modified network, on 
the other hand, the increase is smaller: around 30\% for the T4 spectrum and 50\% for the T5 spectrum. Note that these
values were computing using the case A hydrogen recombination rate, which is the default choice in {\sc enzo}. If we
instead use the case B rate, which is more physically appropriate in the present case, we recover values of $J_{\rm crit}$
that are almost a factor of two larger than those in Table~\ref{tab:compare}. 
  
The highest value quoted in the literature for $J_{\rm crit}$ in the case of a T4 spectrum and a one-zone calculation 
comes from \citet{sbh10}, who find that $J_{\rm crit} = 39$, roughly a factor of two larger than most other estimates.
However, they use the {\sc enzo} chemistry network in their study, and our results here suggest that much of this
factor of two differences is due to the composition of their network and their choice of reaction rate coefficients. It
therefore seems plausible that most of the scatter in the reported values of $J_{\rm crit}$ for the T4 spectrum comes
from differences in the chemical models used for the different studies.
  
\subsubsection{The {\sc krome} package}  
As already noted in Paper I, the {\sc krome} package provides a number of different networks suitable for simulating 
primordial gas, but provides little guidance regarding which network should be used for which application. For
studying direct collapse, the most suitable choice appears to be the {\em react\_xrays} network. We therefore use
this as the basis of our comparison.\footnote{Specifically, we consider the version 
%of the network 
present in the main
{\sc krome} repository on March 26th, 2015, at which time the most recent commit was \texttt{cf3d6b882d35545657d0a5daa7965b5c62c7e970}}
As in the case of the {\sc enzo} network, this omits the effects of the collisional ionization of H by H, and so when
carrying out our comparison, we consider both the unmodified version of the  {\em react\_xrays} network plus a 
modified version that includes this reaction. 

Our results are shown in Table~\ref{tab:compare}. We see that the unmodified network underestimates $J_{\rm crit}$
by around a factor of two. Including the collisional ionization of H by H improves the situation slightly, but even in
this case the {\sc krome} network still significantly underestimates $J_{\rm crit}$. We have investigated the causes
of this discrepancy and find that it is largely driven by the choice of H$^{+}$ recombination rate. The {\sc krome}
 {\em react\_xrays} network adopts the case A rate for this process, which causes the H$^{+}$ abundance to fall off
 more rapidly with increasing density than in our models. If we instead use the more physically appropriate case
 B rate, we find much better agreement between the {\sc krome} results and our own determination of $J_{\rm crit}$.
 This demonstrates that the error introduced into $J_{\rm crit}$ is solely a consequence of the particular choice
 of network and rate coefficients, and not any intrinsic problem within the {\sc krome} infrastructure itself -- if {\sc
 krome} is used simply as an ODE solver and cooling function, with a user-supplied chemical network that includes
 all of the important chemical reactions and an up-to-date set of rate coefficients, then we expect it to be able to
 produce exactly the same values of $J_{\rm crit}$ as we find in our study.
 
\section{Summary}
\label{conc}
In this paper, we have attempted to determine the extent to which estimates of $J_{\rm crit}$ are affected by uncertainties
in the chemical rate coefficients and cooling rates used to model the chemical and thermal evolution of primordial gas.
To do this, we first examined how sensitive $J_{\rm crit}$ is to variation in the rate coefficients of the 26 reactions included
in the reduced chemical network of \citet{glover15}. For those reactions where the sensitivity is large, we then critically examined the
information available regarding the values of their rate coefficients, and quantified the effect of uncertainties in these values on
our derived value of $J_{\rm crit}$. We also carried out a similar analysis for the main cooling processes included in our model.

We found that there are five key chemical reactions where the combination of sensitivity and uncertainty leads to an appreciable
($> 20$\%) uncertainty in $J_{\rm crit}$. These are the collisional dissociation of H$_{2}$ by H (reaction 2), the formation of H$_{2}$
by associative detachment of H$^{-}$ (reaction 3), the formation of H$^{-}$ by radiative association (reaction 6), the collisional 
ionization of H by H (reaction 10) and the collisional detachment of H$^{-}$ by H (reaction 11), although the last of these is important
only in situations where H$^{-}$ photodetachment is unimportant. Amongst these reactions, the single largest uncertainty comes from
the collisional ionization of H by H, a process which is very poorly constrained at the low energies relevant for this study \citep{barklem07}.
In addition, in gas illuminated with a hard T5 spectrum, $J_{\rm crit}$ is also highly sensitive to the value of the H$_{2}$ photodissociation
rate, although in this case the intrinsic uncertainty in the rate is small, and the main difficulty comes from the fact that computationally
efficient treatments of H$_{2}$ self-shielding are often highly over-simplified \citep[see e.g.][]{wh11}.
The total uncertainty introduced into $J_{\rm crit}$ by uncertainties in the remaining 20 reactions is small in comparison to the
effects of these six reactions. Similarly, the remaining uncertainties in the cooling rates of H$_{2}$ and atomic hydrogen do not 
significantly affect $J_{\rm crit}$.

In the unlikely event that the errors in the different rate coefficients combine so as to maximum or minimize $J_{\rm crit}$, we
find that for a T4 spectrum, $J_{\rm crit, min} = 9.8$ and $J_{\rm crit, max} = 47.5$, while for a T5 spectrum, $J_{\rm crit, min}
= 850$ and $J_{\rm crit, max} = 5540$. Therefore, in the worst case, chemical rate coefficient uncertainties could introduce an
uncertainty of around a factor of five into our estimates of $J_{\rm crit}$. In the more likely case that the errors in the different
rate coefficients are uncorrelated, we find that $J_{\rm crit}$ probably lies in the range $J_{\rm crit} = 21 \pm 5$
for a T4 spectrum and $J_{\rm crit} = 2400 \pm 700$ for a T5 spectrum.

Finally, we have compared the values of $J_{\rm crit}$ that we have derived using our reaction network and rate coefficients
with those that we obtain if we use instead the network and rate coefficients from (a) the {\sc enzo} hydrodynamical code or
(b) the {\sc krome} astrochemistry package, both of which have been used in a number of previous studies of the direct
collapse model. We find that the {\sc enzo} chemical model tends to overestimate $J_{\rm crit}$ by around a factor of two,
although this discrepancy increases to a factor of four if the physically appropriate case B hydrogen recombination rate is
used in place of the case A rate that is the default in {\sc enzo}. The {\sc krome} model, on the other hand, underestimates
$J_{\rm crit}$ by around a factor of two.

\section*{Acknowledgements}
The author would like to thank the anonymous referee for their detailed report on an earlier version of this paper, 
and for calculating the rate coefficient for H-H collisional ionization quoted in Equation~\ref{ksrate}. The author
would also like to thank B.~Agarwal and B.~Smith for prompting him to think about this problem in the first place.
Financial support for this work was provided by the Deutsche Forschungsgemeinschaft  via
SFB 881, ``The Milky Way System'' (sub-projects B1, B2 and B8) and SPP 1573, ``Physics of the Interstellar Medium'' (grant number GL 668/2-1), 
and by the European Research Council under the European Community's Seventh Framework Programme (FP7/2007-2013) via the 
ERC Advanced Grant STARLIGHT (project number 339177).


\begin{thebibliography}{}

\bibitem[Abel et~al.(1997)]{abel97}
Abel, T., Anninos, P., Zhang, Y., \& Norman, M.~L.\ 1997, New Astron., 2, 181

\bibitem[Abgrall \& Roueff(1989)]{ar89}
Abgrall, H., \& Roueff, E.\ 1989, A\&AS, 79, 313

\bibitem[Agarwal et~al.(2012)]{agar12}
Agarwal, B., Khochfar, S., Johnson, J.~L., Neistein, E., {Dalla Vecchia}, C., \& Livio, M.\ 2012, MNRAS, 425, 2854

\bibitem[Agarwal \& Khochfar(2015)]{ak15}
Agarwal, B., \& Khochfar, S.\ 2015, MNRAS, 446, 160

\bibitem[Agarwal et~al.(2015)]{agar15}
Agarwal, B., Smith, B., Glover, S.~C.~O., Khochfar, S., \& Natarajan, P., 2015, MNRAS, in prep.\

\bibitem[Ahn et~al.(2009)]{ahn09}
Ahn, K., Shapiro, P.~R., Iliev, I.~T., Mellema, G., \& Pen, U.-L.\ 2009, ApJ, 695, 1430

\bibitem[Bardsley, Herzenberg \& Mandl(1966a)]{bard66a}
Bardsley, J.~N., Herzenberg, A., \& Mandl, F.\ 1966a, Proc.\ Phys.\ Soc., 89, 305

\bibitem[Bardsley, Herzenberg \& Mandl(1966b)]{bard66b}
Bardsley, J.~N., Herzenberg, A., \& Mandl, F.\ 1966b, Proc.\ Phys.\ Soc., 89, 321

\bibitem[Barklem(2007)]{barklem07}
Barklem, P.~S.\ 2007, A\&A, 466, 327

\bibitem[Barklem et~al.(2011)]{barklem11}
Barklem, P.~S., Belyaev, A.~K., Guitou, M., Feautrier, N., Gad\'ea, F.~X., \&
Spielfiedel, A.\ 2011, A\&A, 530, A94

\bibitem[Begelman, Volonteri, \& Rees(2006)]{begel06}
Begelman, M.~C., Volonteri, M., \& Rees, M.~J.\ 2006, MNRAS, 370, 289

\bibitem[Begelman(2010)]{begel10}
Begelman, M.~C.\ 2010, MNRAS, 402, 673

\bibitem[Black(1981)]{black81}
Black, J.~H.\ 1981, MNRAS, 197, 553

\bibitem[Boothroyd et~al.(1991)]{bkmp1}
Boothroyd, A.~I., Martin, P.~G., Keogh, W.~J., \& Peterson, M.~R.\ 1991, J.\ Chem.\ Phys., 95, 4343

\bibitem[Boothroyd et~al.(1996)]{bkmp2}
Boothroyd, A.~I., Keogh, W.~J., Martin, P.~G., \& Peterson, M.~R.\ 1996, J.\ Chem.\ Phys., 104, 7139

\bibitem[Bray \& Stelbovics(1993)]{bs93}
Bray, I., \& Stelbovics, A.~T.\ 1993, Phys.\ Rev.\ Lett., 70, 746

\bibitem[Bromm \& Loeb(2003)]{bl03}
Bromm, V., \& Loeb, A.\ 2003, ApJ, 596, 34

\bibitem[Browne \& Dalgarno(1969)]{bd69}
Browne, J.~C., \& Dalgarno, A.\ 1969, J.\ Phys.\ B, 2, 885

\bibitem[Bryan et~al.(2014)]{bryan14}
Bryan, G.~L., et~al.\ 2014, ApJS, 211, 19

%\bibitem[Bruhns et al.(2010)]{bru10}
%Bruhns, H., Kreckel, H., Miller, K.~A., Urbain, X., \& Savin, D.~W.\ 2010, Phys.\ Rev.\ A, 
%82, 042708

\bibitem[Cen(1992)]{cen92}
Cen, R.\ 1992, ApJS, 78, 341

\bibitem[Chen \& Peacher(1968)]{cp68}
Chen, J.~C.~Y., \& Peacher, J.~L.\ 1968, Phys.\ Rev., 167, 30

\bibitem[C\'izek et al.(1998)]{cizek98}
C\'izek, M., Hor\'acek, J., \& Domcke, W.\ 1998, J.\ Phys.\ B, 31, 2571

\bibitem[Clark, Glover \& Klessen(2012)]{cgk12}
Clark, P.~C., Glover, S.~C.~O., \& Klessen, R.~S.\ 2012, MNRAS, 420, 745

\bibitem[Coppola et al.(2011)]{cop11}
Coppola, C.~M., Longo, S., Capitelli, M., Palla, F., \& Galli, D.\ 2011,
ApJS, 193, 7

\bibitem[Croft, Dickinson \& Gadea(1999)]{cdg99}
Croft, H., Dickinson, A.~S., \& Gadea, F.~X.\ 1999, MNRAS, 304, 327

\bibitem[Dalgarno \& Lepp(1987)]{dl87}
Dalgarno, A., \& Lepp, S.\ 1987, in Astrochemistry, ed.\ M.~S.\ Vardya
\& S.~P.\ Tarafdar (Dordrecht: Reidel), 109

\bibitem[{de Jong}(1972)]{dejong72}
{de Jong}, T.\ 1972, A\&A, 20, 263

\bibitem[Dijkstra et~al.(2008)]{dijk08}
Dijkstra, M., Haiman, Z., Mesinger, A., Wyithe, J.~S.~B.\ 2008, MNRAS, 391, 1961

%\bibitem[Doughty \& Fraser(1964)]{df64}
%Doughty, N.~A., \& Fraser, P.~A., 1964, Atomic Collision Processes (Amsterdam: North-Holland)

\bibitem[Draine \& Bertoldi(1996)]{db96}
Draine, B.~T., \& Bertoldi, F.\ 1996, ApJ, 468, 269

\bibitem[Drawin(1968)]{drawin68}
Drawin, H.-W., 1968, Zeit.\ Phys.\ 211, 404

\bibitem[Drawin(1969)]{drawin69}
Drawin, H.-W., 1969, Zeit.\ Phys.\ 225, 470

\bibitem[Esposito \& Capitelli(2009)]{ec09}
Esposito, F., \& Capitelli, M.\ 2009, J.\ Phys.\ Chem.\ A, 113, 15307

\bibitem[Ferland et~al.(1992)]{fer92}
Ferland, G.~J., Peterson, B.~M., Horne, K., Welsh, W.~F., \& Nahar, S.~N.\ 1992, ApJ, 387, 95

\bibitem[Forrey et~al.(1997)]{forrey97}
Forrey, R.~C., Balakrishnan, N., Dalgarno, A., \& Lepp, S.\ 1997, ApJ, 489, 1000

\bibitem[Fussen \& Kubach(1986)]{fk86}
Fussen, D., \& Kubach, C.\ 1986, J.\ Phys.\ B, 19, L31

\bibitem[Galli \& Palla(1998)]{gp98}
Galli, D., \& Palla, F., 1998, A\&A, 335, 403

\bibitem[Gealy \& {van Zyl}(1987)]{gvz87}
Gealy, M.~W., \& {van Zyl}, B.\ 1987, Phys.\ Rev.\ A, 36, 3100

\bibitem[Gerlich et~al.(2012)]{ger12}
Gerlich, D., Jusko, P., Rou\v{c}ka, \v{S}., Zymak, I., Pla\v{s}il, R., \&
Glos\'{i}k, J.\ 2012, ApJ, 749, 22

\bibitem[Glover(2011)]{glover11}
Glover, S.~C.~O., 2011, Habilitation thesis, Univ.\ Heidelberg

\bibitem[Glover(2015)]{glover15}
Glover, S.~C.~O., 2015, MNRAS, 451, 2082 (Paper I)

\bibitem[Glover \& Abel(2008)]{ga08}
Glover, S.~C.~O., \& Abel, T.\ 2008, MNRAS, 388, 1627

\bibitem[Glover, Savin \& Jappsen(2006)]{gsj06}
Glover, S.~C., Savin, D.~W., \& Jappsen, A.-K.\ 2006, ApJ, 640, 553

\bibitem[Glover \& Savin(2009)]{Glover09}
Glover, S.~C.~O., \& Savin, D.~W.\ 2009, MNRAS, 393, 911

\bibitem[Grassi et~al.(2014)]{grassi14}
Grassi, T., Bovino, S., Schleicher, D.~R.~G.,  Prieto, J., Seifried, D., Simoncini, E., \& Gianturco, F.~A.\ 2014, MNRAS, 439, 2386

\bibitem[Greif \& Bromm(2006)]{gb06}
Greif, T.~H., \& Bromm, V.\ 2006, MNRAS, 373, 128

\bibitem[Grosser \& Kr\"uger(1984)]{gk84}
Grosser, J., \& Kr\"uger, W.\ 1984, Z.\ Phys.\ A, 318, 25

\bibitem[Gryzinski(1965)]{gryz65}
Gryzinski, M., 1965, Phys.\ Rev.\ A, 138, 336

\bibitem[Haiman, Abel \& Rees(2000)]{har00}
Haiman, Z., Abel, T., \& Rees, M.~J.\ 2000, ApJ, 534, 11

\bibitem[Hartwig et~al.(2015)]{hartwig15}
Hartwig, T., Glover, S.~C.~O., Klessen, R.~S., Latif, M.~A., \& Volonteri, M.\ 2015, MNRAS, 452, 1233

\bibitem[Hill, Geddes \& Gilbody(1979)]{hgg79}
Hill, J., Geddes, J., \& Gilbody, H.~B.\ 1979, J.\ Phys.\ B, 12, 3341

\bibitem[Hollenbach \& McKee(1979)]{hm79}
Hollenbach, D., \& McKee, C.~F.\ 1979, ApJS, 41, 555

\bibitem[Hollenbach \& McKee(1989)]{hm89}
Hollenbach, D., \& McKee, C.~F.\ 1989, ApJ, 342, 306

\bibitem[Hui \& Gnedin(1997)]{hg97}
Hui, L., \& Gnedin, N.~Y.\ 1997, MNRAS, 292, 27

\bibitem[Hutchins(1976)]{hutch76}
Hutchins, J.~B., 1976, ApJ, 205, 103

\bibitem[Inayoshi \& Omukai(2011)]{io11}
Inayoshi, K., \& Omukai, K.\ 2011, MNRAS, 416, 2748

\bibitem[Inayoshi \& Tanaka(2015)]{it14}
Inayoshi, K., \& Tanaka, T.~L.\ 2015, MNRAS, 450, 4350

\bibitem[Janev, Langer \& Evans(1987)]{janev87}
Janev, R.~K., Langer, W.~D., \& Evans, K.\ 1987, Elementary Processes in Hydrogen-Helium plasmas (Berlin: Springer)

\bibitem[Janev, Reiter \& Samm(2003)]{janev03}
Janev, R.~K., Reiter, D., \& Samm, U.\ 2003, Berichte des Forschungszentrums J\"ulich, 4105

\bibitem[Kreckel et~al.(2010)]{kreck10}
Kreckel, H., Bruhns, H., \v{C}\'i\v{z}ek, M., Glover, S.~C.~O., Miller, K.~A., 
Urbain, X., \& Savin, D.~W.\ 2010, Science, 329, 69

\bibitem[Kunc \& Soon(1991)]{ks91}
Kunc, J.~A., \& Soon, W.~H.\ 1991, J.\ Chem.\ Phys., 95. 5738

\bibitem[Latif et~al.(2014)]{latif14}
Latif, M.~A., Bovino, S., {Van Borm}, C., Grassi, T., Schleicher, D.~R.~G., \& Spaans, M.\
2014, MNRAS, 443, 1979

\bibitem[Latif et~al.(2015)]{latif15}
Latif, M.~A., Bovino, S., Grassi, T., Schleicher, D.~R.~G., \& Spaans, M.\ 2015, MNRAS,  446, 3163

\bibitem[Launay et~al.(1991)]{lau91}
Launay, J.~M., {Le Dourneuf}, M., \& Zeippen, C.~J.\ 1991, A\&A, 252, 842

\bibitem[Lenzuni, Chernoff \& Salpeter(1991)]{lcs91}
Lenzuni, P., Chernoff, D.~F., \& Salpeter, E.~E.\ 1991, ApJS, 76, 759

\bibitem[Lepp, Buch \& Dalgarno(1995)]{lbd95}
Lepp, S., Buch, V., \& Dalgarno, A.\ 1995, ApJS, 98, 345

\bibitem[Liu(1973)]{l73}
Liu, B.\ 1973, J.\ Chem.\ Phys., 58, 1925

\bibitem[Lotz(1967)]{lotz67}
Lotz, W.\ 1967, ApJS, 14, 207

\bibitem[Mandy \& Martin(1993)]{mm93}
Mandy, M.~E., \& Martin, P.~G.\ 1993, ApJS, 86, 199

\bibitem[Martin, Keogh \& Mandy(1998)]{mkm98}
Martin, P.~G., Keogh, W.~J., \& Mandy, M.~E.\ 1998, ApJ, 499, 793

\bibitem[Martin, Schwarz \& Mandy(1996)]{msm96}
Martin, P.~G., Schwarz, D.~H., \& Mandy, M.~E.\ 1996, ApJ, 461, 265

%\bibitem[Miller et~al.(2011)]{miller11}
%Miller, K.~A., Bruhns, H., Eli\'a\v{s}ek, J., \v{C}\'i\v{z}ek, M., Kreckel, H., Urbain, X., \& Savin, D.~W.\
%2011, Phys.\ Rev.\ A, 84, 052709

\bibitem[McClure(1968)]{mcc68}
McClure, G.~W.\ 1968, Phys.\ Rev., 166, 22

\bibitem[Mielke, Garrett \& Peterson(2002)]{mgp02}
Mielke, S.~L., Garrett, B.~C., \& Peterson, K.~A.\ 2002, J.\ Chem.\ Phys., 116, 4142

\bibitem[Miyake et~al.(2010)]{miyake10}
Miyake, S., Stancil, P.~C., Sadeghpour, H.~R., Dalgarno, A., McLaughlin, B.~M., \& Forrey, R.~C.\ 2010, ApJ, 709, L168

\bibitem[Moseley et~al.(1970)]{map70}
Moseley, J., Aberth, W., \& Peterson, J.~R.\ 1970, Phys.\ Rev.\ Lett.\ 24, 435

\bibitem[Omukai(2001)]{om01}
Omukai, K., 2001, ApJ, 546, 635

\bibitem[Ramaker \& Peek(1976)]{rp76}
Ramaker, D.~E., \& Peek, J.~M.\ 1976, Phys.\ Rev.\ A, 13, 58

\bibitem[Riley \& Burke(1999)]{rb99}
Riley, M.~E., \& Burke, R.~A.\ 1999, J.\ Phys., B, 32, 5279

\bibitem[Schleicher et~al.(2013)]{schleicher13}
Schleicher, D.~R.~G., Palla, F., Galli, D., \& Latif, M.\ 2013, A\&A, 558, A59

\bibitem[Schleicher, Spaans \& Glover(2010)]{ssg10}
Schleicher, D.~R.~G., Spaans, M., \& Glover, S.~C.~O.\ 2010, ApJ, 712, L69

\bibitem[Scholz \& Walters(1991)]{sw91}
Scholz, T.~T., \& Walters, H.~R.~J.\ 1991, ApJ, 380, 302

\bibitem[Shah, Elliott \& Gilbody(1987)]{seg87}
Shah, M.~B., Elliott, D.~S., \& Gilbody, H.~B.\ 1987, J.\ Phys.\ B, 20, 3501

\bibitem[Shang, Bryan \& Haiman(2010)]{sbh10}
Shang, C., Bryan, G.~L., \& Haiman, Z.\ 2010, MNRAS, 402, 1249

\bibitem[Shapiro \& Kang(1987)]{sk87}
Shapiro, P.~R., \& Kang, H.\ 1987, ApJ, 318, 32

\bibitem[Shingal, Bransden \& Flower(1989)]{sbf89}
Shingal, R., Bransden, B.~H., \& Flower, D.~R.\ 1989, J.\ Phys., B, 22, 855

\bibitem[Shull(1978)]{shull78}
Shull, J.~M.\ 1978, ApJ, 219, 877

\bibitem[Siegbahn \& Liu(1978)]{sl78}
Siegbahn, P., \& Liu, B.\ 1978, J.\ Chem.\ Phys., 68, 2457

\bibitem[Soon(1992)]{soon92}
Soon, W.~H.\ 1992, ApJ, 394, 717

\bibitem[Stancil, Babb \& Dalgarno(1993)]{sbd93}
Stancil, P.~C., Babb, J.~F., \& Dalgarno, A.\ 1993, ApJ, 414, 672

\bibitem[Stancil, Lepp \& Dalgarno(1998)]{sld98}
Stancil, P.~C., Lepp, S., \& Dalgarno, A.\ 1998, ApJ, 509, 1

\bibitem[Stenrup, Larson \& Elander(2009)]{sle09}
Stenrup, M., Larson, A., \& Elander, N.\ 2009, Phys.\ Rev.\ A, 79, 012713

\bibitem[Sugimura, Omukai \& Inoue(2014)]{soi14}
Sugimura, K., Omukai, K., \& Inoue, A.~K.\ 2014, MNRAS, 445, 544

\bibitem[Sun \& Dalgarno(1994)]{sd94}
Sun, Y., \& Dalgarno, A.\ 1994, ApJ, 427, 1053

\bibitem[Tanaka \& Haiman(2009)]{th09}
Tanaka, T., \& Haiman, Z.\ 2009, ApJ, 696, 1798

\bibitem[Truhlar \& Horowitz(1978)]{th78}
Truhlar, D.~G., \& Horowitz, C.~J.\ 1978, J.\ Chem.\ Phys., 68, 2466

\bibitem[Urbain et~al.(2012)]{urbain12}
Urbain, X., Lecointre, J., Mezdari, F., Miller, K.~A., \& Savin, D.~W.\ 2012, J.\ Phys.\ Conf.\ Ser.,
388, 092004

\bibitem[Varandas et~al.(1987)]{dmbe}
Varandas, A.~J.~C., Brown, F.~B., Mead, C.~A., Truhlar, D.~G., \& Blais, N.~C.\ 1987, J.\ Chem.\ Phys., 86, 6258

\bibitem[Wishart(1979)]{wish79}
Wishart, A.~W., 1979, MNRAS, 187, P59

\bibitem[Wolcott-Green \& Haiman(2011)]{wh11}
Wolcott-Green, J., \& Haiman, Z.\ 2011, MNRAS, 412, 2603

\bibitem[Wolcott-Green, Haiman \& Bryan(2011)]{whb11}
Wolcott-Green, J., Haiman, Z., \& Bryan, G.~L.\ 2011, MNRAS, 418, 838

\bibitem[Wolniewicz, Simbotin \& Dalgarno(1998)]{wol98}
Wolniewicz, L., Simbotin, I., \& Dalgarno, A.\ 1998, ApJS, 115, 293

\bibitem[Wrathmall \& Flower(2007)]{wf07}
Wrathmall, S.~A., \& Flower, D.~R.\ 2007, J.\ Phys.\ B, 40, 3221

\bibitem[Wrathmall, Gusdorf \& Flower(2007)]{wgf07}
Wrathmall, S.~A., Gusdorf, A., \& Flower, D.~R.\ 2007, MNRAS, 382, 133

\end{thebibliography}
\end{document}